\documentclass[12pt]{article}
\usepackage{jheppub}
\usepackage{subfigure}
\usepackage{amsmath,amssymb,bm,graphicx,subfigure,makecell}
\usepackage{lineno,booktabs,url,color}
\usepackage[numbers,sort&compress]{natbib}

\title{A global analysis of Energy-Energy Correlation data: determination of $\bm{\alpha_S}$ and non-perturbative QCD parameters}

\author[a,b]{Ugo Giuseppe Aglietti}
\author[c,d]{Giancarlo Ferrera,}
\author[c,d]{Lorenzo Rossi}

\affiliation[a]{Dipartimento di Fisica, Universit\`a di Roma `La Sapienza', I-00185 Rome, Italy}
\affiliation[b]{ INFN, Sezione di Roma,
I-00185 Rome, Italy}
\affiliation[c]{Dipartimento di Fisica, Università di Milano, Via G. Celoria 16, I-20133 Milan, Italy}
\affiliation[d]{INFN, Sezione di Milano, Via G. Celoria 16, I-20133 Milan, Italy}

\abstract{We present a comprehensive global analysis of Energy-Energy Correlation (EEC) data in
electron-positron annihilation into hadrons, spanning a wide range of center-of-mass energies
($7.7\,\,\text{GeV}\!\leq\!\sqrt{s}\!\leq\! 91.2\,\,\text{GeV})$.
In the back-to-back (two-jet) region, we resum to all orders the logarithmically-enhanced contributions up to next-to-next-to-next-to-leading logarithmic (N$^3$LL) accuracy. The resummed results are consistently matched to fixed-order calculations up to $\mathcal{O}(\alpha_S^3)$. 
Our resummation formalism also incorporates dominant heavy-quark mass effects and 
models non-perturbative power corrections
by means of an analytic dispersive approach.
A simultaneous fit yields an excellent description of experimental data across all energies, 
enabling a precise determination of the strong coupling, $\alpha_S(m_Z^2) = 0.1192 \pm 0.0028$,  
as well as the non-perturbative parameters, including those characterizing the Collins--Soper  evolution kernel.
Our analysis includes, for the first time in a global fit, 
datasets from the ALEPH and AMY collaborations.}

\emailAdd{ugo.aglietti@roma1.infn.it}
\emailAdd{giancarlo.ferrera@mi.infn.it}
\emailAdd{lorenzo.rossi3@unimi.it}

\usepackage{hyperref}
\begin{document}
\maketitle

\section{Introduction}

The study of the Energy-Energy Correlation (EEC) function 
in high-energy electron-positron annihilation~\cite{Basham:1978zq,Basham:1978bw} 
remains a cornerstone for testing the perturbative regime of Quantum Chromodynamics (QCD), 
for testing non-perturbative (NP) hadronization models, and for the precise extraction of the strong coupling constant $\alpha_S$ at a reference energy scale~\cite{Moult:2025nhu}.

The EEC describes the distribution of the angular separation $\chi$ between pairs of energetic particles in the final state. 
Perturbative QCD calculations of the EEC, based on a fixed-order truncation in $\alpha_S$, were made at leading order (LO, $\mathcal{O}(\alpha_S)$) in the late 1970's~\cite{Basham:1978zq}. Subsequently, next-to-leading order 
(NLO, $\mathcal{O}(\alpha_S^2)$) corrections were obtained both numerically and analytically~\cite{Richards:1982te,Richards:1983sr,Kunszt:1989km,Dixon:2018qgp}. So far, next-to-next-to-leading order (NNLO, $\mathcal{O}(\alpha_S^3)$) corrections have been determined only numerically, through Monte Carlo integrations of the fully differential cross section for three-jet production in $e^+e^-$ annihilation at NNLO~\cite{Tulipant:2017ybb,DelDuca:2016csb,DelDuca:2016ily}.

In the region where the final-state hadrons 
are mostly emitted in two nearly back-to-back directions ($\chi \to \pi$), the fixed-order QCD perturbative expansion becomes unreliable, since its coefficients are  enhanced by Sudakov logarithms, namely large double logarithms of infrared origin. The systematic resummation of these logarithmic contributions to all perturbative orders is crucial in order to obtain reliable predictions in 
this region. The resummation of EEC in the back-to-back region has been achieved at the next-to-leading logarithmic (NLL) accuracy in Refs.\,\cite{Collins:1981uk,Collins:1981va,Collins:1985xx,Kodaira:1981nh,Kodaira:1982az} and next-to-next-to-leading logarithmic (NNLL) accuracy in Refs.\,\cite{deFlorian:2004mp,Tulipant:2017ybb,Kardos:2018kqj}. More recently, resummed predictions at the  next-to-next-to-next-to-leading logarithmic accuracy (N$^3$LL) have been obtained both in the framework of Soft-Collinear Effective Theory (SCET)~\cite{Moult:2018jzp,Ebert:2020sfi,Jaarsma:2025tck}
and within full QCD~\cite{Aglietti:2024xwv}. 

The perturbative expansion of the EEC function also contains large (single) infrared logarithmic corrections in the 
forward region ($\chi \to 0^+$), where two or more energetic
hadrons are produced at small angular separations
(the so-called intra-jet activity). However, these effects are physically
quite different from the two-jet ones, being of hard-collinear nature~\cite{Aglietti:2024xwv}.
Finally, a realistic description of the 
(experimental, i.e.\ hadronic) EEC spectrum requires the inclusion of power-suppressed non-perturbative QCD contributions~\cite{Kodaira:1981nh,Fiore:1992sa,Nason:1995np,Dokshitzer:1995qm,Dokshitzer:1999sh,Aglietti:2024xwv}.

In Ref.\,\cite{Aglietti:2024xwv}, 
theoretical spectra, involving the resummation of Sudakov logarithms in the back-to-back region up to N$^3$LL, matched
up to NNLO, have been compared with experimental data at the $Z$-boson resonance. The present study extends the results of Ref.\,\cite{Aglietti:2024xwv}, by means of a global fit to all available datasets at different energies, and extracts both the strong coupling constant at a reference scale, as well as the non-perturbative QCD parameters.  
We perform the fit by consistently
including data points from both the peak and the intermediate regions. This choice ensures the numerical stability of the fit and a reliable separation between perturbative and non-perturbative dynamics.
Moreover, we show that the inclusion of  non-perturbative power-behaved QCD effects
through the analytic ``dispersive approach" of Refs.\,\cite{Aglietti:2024xwv}, 
together with a suitable 
energy-dependent non-perturbative evolution, provides an excellent description of the data across all energies.

We believe that the use of an analytic dispersive model for non-perturbative power corrections is a significant advantage of our method; Unlike methods which extract hadronization corrections by means of general-purpose 
Monte-Carlo event generators
(which may introduce tune-dependent biases and unclear theoretical uncertainties), our approach treats 
non-perturbative effects via an analytic parameterization, which is fitted simultaneously with $\alpha_S$. 
This ensures better control over both QCD perturbative and non-perturbative effects.

In the end, we obtain a precise determination of $\alpha_S(m_Z^2)=0.1192\pm 0.0028$, which turns out to be fully consistent with the current global average~\cite{ParticleDataGroup:2024cfk}. 
Moreover, by analyzing EEC spectra at
different energies simultaneously, we disentangle the energy-dependent components of the non-perturbative form factor, enabling a novel extraction of the Collins--Soper kernel.

A novel aspect of our global analysis is the inclusion of very recent, high-precision measurements of the EEC from the Electron-Positron Alliance collaboration \cite{Electron-PositronAlliance:2025wzh,Electron-PositronAlliance:2025fhk}, based on a reanalysis of archived ALEPH data available at $\sqrt{s}=91.2\,\text{GeV}$. 
Furthermore, we extend the experimental coverage of our global analysis by including previously unused datasets from the AMY collaboration at 
$\sqrt{s} = 58.0$ GeV allowing for a more robust test of the QCD evolution predicted by the Collins-Soper kernel.
We will show that our theoretical framework successfully achieves an excellent description of these 
precise measurements alongside the legacy data.

%\LRcom{A novel aspect of our global analysis is the inclusion of the unused datasets from the AMY collaboration~\cite{AMY:1997waz} at  $\sqrt{s} = 58.0$ GeV allowing for a more robust test of the QCD evolution predicted by the Collins-Soper kernel and also a comparison with very recent, high-precision measurements of the EEC from the Electron-Positron Alliance collaboration \cite{Electron-PositronAlliance:2025wzh,Electron-PositronAlliance:2025fhk}, based on a novel reanalysis of archived ALEPH data available at $\sqrt{s}=91.2\,\text{GeV}$, which tests the predictive power of QCD.}

The structure of the paper is as follows. In Sec.\,\ref{s:formalism} we present a concise review of the theoretical framework. Sec.\,\ref{s:analysis} describes the datasets, the treatment of the uncertainties and the methodology used for the parameter extraction. The results, including a brief discussion of heavy quark mass effects, are presented in Sec.\,\ref{s:results}. Our conclusions are summarized in Sec.\,\ref{s:conclusions}.

\section{Formalism}
\label{s:formalism}

The Energy-Energy Correlation (EEC) function is defined as:
\begin{equation}
\label{e:EEC_def}
\frac{d\Sigma}{d\cos\chi}
= \sum_{n=2}^\infty \sum_{i,j=1}^n \int
\frac{E_i}{Q}\,\frac{E_j}{Q}\,
\delta(\cos\chi - \cos\theta_{ij})\,
d\sigma^{(n)}_{e^+ e^- \to h_i h_j + X}\, ,
\end{equation}
where the sum runs over all pairs of final-state particles $(i,j)$, with energies $E_i$ and $E_j$, for events containing $n \geq 2$ hadrons. The variable $\theta_{ij}$ denotes the relative angle between their spatial momenta, and the hard scale $Q=\sqrt{s}$ is identified with the center-of-mass energy of the colliding $e^+ e^-$ pair.

It is useful to express the EEC in terms of the dimensionless unitary variable
\begin{equation}
    z \equiv \frac{1-\cos\chi}{2} = \sin^2\frac{\chi}{2}\, .
\end{equation}
Experimental measurements are typically presented after normalization to the total cross section $\sigma_{\rm tot}$, defined as
\begin{equation}
    \sigma_{tot} = \int_{-1}^{+1} \frac{d\Sigma}{d\cos \chi} d\cos \chi = \int_0^1 \frac{d\Sigma}{dz} dz = \int \Bigg{(} \sum_{i=1}^n  \frac{E_i}{Q}\Bigg{)}^2 d\sigma  = \int d\sigma\, . 
\end{equation}

\subsection{Resummation in the two-jet region}

In the back-to-back region ($z \to 1^-$ or $\chi \to \pi$), in which all energetic hadrons in the final states are collimated in two opposite
directions, the convergence of the standard fixed-order perturbative QCD expansion is spoiled, since its coefficients 
are enhanced by large double logarithms of infrared (soft and collinear) origin. To obtain
reliable perturbative predictions, these so-called Sudakov logarithms have to be resummed to all orders in 
$\alpha_S$. 
For this purpose, the normalized EEC distribution is conveniently decomposed as
\begin{equation}
    \frac{1}{\sigma_{tot}} \frac{d\Sigma}{d z} =     \frac{1}{\sigma_{tot}} \frac{d\Sigma_{(res.)}}{d z} +  \frac{1}{\sigma_{tot}} \frac{d\Sigma_{(fin.)}}{d z}\, , 
    \label{e:dsdz}
\end{equation}
where the first term on the r.h.s., the {\itshape resummed} term, factorizes and resums to all orders the logarithmically enhanced contributions (it includes also the $\delta(1-z)$ terms), while the second term denotes the {\itshape finite} 
(i.e.\ free from large infrared logarithms) remainder function and is computed at fixed order.

In order to consistently take into account the kinematic constraint of transverse momentum conservation in multiple parton emissions, the resummation is made in the impact-parameter $b$-space~\cite{Collins:1981uk,Collins:1981va,Kodaira:1981nh}, where $b$ is conjugate to the variable $Q\sqrt{1-z}$.
In $b$-space, the back-to-back limit $1-z \ll 1$ corresponds to the region $bQ \gg 1$
and the large Sudakov logarithms are of the form $\alpha_S^n \ln^k(Q^2 b^2)$, with $1 \leq k \leq 2n$. 
After the resummation procedure has been carried out, the resummed part of the distribution in physical space is recovered by means of an inverse Fourier--Bessel transform:
\begin{equation} 
\frac{1}{\sigma_{tot}} \frac{d\Sigma_{(res.)}}{d z} = \frac{1}{2} H(\alpha_S) \int_0^\infty d(Qb) \frac{Qb}{2} J_0(\sqrt{1 -z}\, Qb) \,S(b,Q)\,,
\label{e:resum}
\end{equation}
where $J_0(x)$ denotes the Bessel function of the first kind of zero order. 

As shown in Eq.\,\eqref{e:resum}, the resummed component of the EEC distribution factorizes
into a hard function $H(\alpha_S)$ and a Sudakov form factor $S(b,Q)$. The hard function $H(\alpha_S)$, which 
encodes hard-virtual corrections and is independent of $b$, 
admits a standard perturbative expansion:
\begin{align}
    H(\alpha_S) &= 1 + \sum_{n=1}^\infty \Big{(}\frac{\alpha_S}{\pi}\Big{)}^n \, H_n \,,
\end{align}
where the QCD coupling $\alpha_S\equiv\alpha_S(\mu_R^2)$ is evaluated at the renormalization scale $\mu_R = \mathcal{O}(Q)$.

The Sudakov form factor $S(b,Q)$ resums, to all orders in the conjugate $b$-space, 
the large Sudakov logarithms
and can be expressed in the following exponential form~\cite{Collins:1981uk,Kodaira:1981nh,Bozzi:2005wk}:
\begin{equation}
\label{e:sudakov_q}
    S(b,Q) = \exp \Bigg{\{}  - \int_{b_0^2/b^2}^{\mu_Q^2} \frac{dq^2}{q^2} \Bigg{[} A\Big{(}\alpha_S(q^2)\Big{)}
    \ln
    \frac{Q^2}{q^2}
    + B \Big{(}\alpha_S(q^2)\Big{)}\Bigg{]}\Bigg{\}}\, ,
\end{equation}
where $b_0 \equiv 2 \exp{(-\gamma_E)} \simeq 1.123$ ($\gamma_E = 0.5772\cdots$ is the Euler-Mascheroni number)
is a kinematic coefficient and 
$\mu_Q$ denotes the resummation scale which takes into account the arbitrariness associated in the resummation procedure~\cite{Bozzi:2005wk}. The central (reference) values of $\mu_R$ and $\mu_Q$ have to be chosen  of the order of the hard scale $Q$. Variations around the central values are typically made to estimate the corresponding perturbative uncertainties in the renormalization and resummation procedure.
The perturbative expansion of the double-logarithmic and single-logarithmic functions $A(\alpha_S)$ and $B(\alpha_S)$ in 
Eq.\,\eqref{e:sudakov_q} reads 
\begin{align}
    A(\alpha_S) &= \sum_{n=1}^\infty \Big{(}\frac{\alpha_S}{\pi}\Big{)}^n \, A_n \, , \\ 
    B(\alpha_S) &= \sum_{n=1}^\infty \Big{(}\frac{\alpha_S}{\pi}\Big{)}^n \, B_n \, .
\end{align}
After analytical integration in Eq.\,\eqref{e:sudakov_q}, the Sudakov form factor can be explicitly written in the form~\cite{Catani:2000vq,deFlorian:2004mp,Aglietti:2024xwv}:
\begin{equation}
    S(b,Q) = \exp \Bigg{\{} \widetilde{L}\,g_1(\lambda) + g_2(\lambda) + \frac{\alpha_S}{\pi} g_3(\lambda) + \Big{(}  \frac{\alpha_S}{\pi} \Big{)}^2 g_4(\lambda) + \sum_{n=3}^{+\infty} \Big{(}  \frac{\alpha_S}{\pi} \Big{)}^n g_{n+2}(\lambda) \Bigg{\}},
\end{equation}
with
\begin{equation}
\lambda \equiv \frac{\alpha_S(\mu_R^2)}{\pi} \beta_0 \widetilde{L}\, \quad \text{and}\quad \, \widetilde{L} = \ln \Bigg{(} \frac{\mu_Q^2 b^2}{b_0^2} +1 \Bigg{)}\,,
\end{equation}
where $\beta_0$ is the first-order coefficient of the QCD $\beta$-function.
The use of the logarithmic variable 
$\widetilde{L}$ (rather than
the usual 
$L \equiv \ln (\mu_Q^2 b^2/b_0^2)$)
ensures the correct unitarity constraint~\cite{Bozzi:2005wk} and prevents spurious unphysical divergences in the limit $b\rightarrow0^+$.
The explicit analytic expressions of the functions $g_n(\lambda)$ up to next-to-next-to-next-to-leading logarithmic (N$^3$LL), 
i.e.\ up to $g_4(\lambda)$ included,
together with the numerical values of the corresponding coefficients, can be found in Ref.\,\cite{Aglietti:2024xwv}.

We observe that the functions $g_n(\lambda)$  depend explicitly both on the unphysical renormalization ($\mu_R$)
 and  resummation ($\mu_Q$) scales. 
This artificial dependence reduces by increasing the perturbative accuracy of the calculation
and formally vanishes by including all
orders.
We emphasize that, due to the singularity  of the QCD coupling at the scale $\Lambda_{QCD}$
(a simple pole at lowest order), the functions $g_n(\lambda)$ are singular at the point $\lambda=1$, which actually corresponds to 
the value of the impact parameter $b_L = b_0/Q \exp[\pi/(2\beta_0 \alpha_S)] \simeq 1/\Lambda_{QCD}$. 
Therefore, in order to evaluate the integral in Eq.\,\eqref{e:resum}, a non-perturbative prescription is needed.
In this work, we use the so-called $b_\star$ prescription~\cite{Collins:1984kg,Collins:1981va,Collins:2011zzd,Collins:1985xx}, which regularizes the integral by freezing the value of the impact parameter $b$ before it reaches the Landau singularity at $b_L$. 
This is achieved by introducing a parameter $b_{\text{max}}<b_L$ and replacing $b$ as
\begin{equation}
\label{e:bstar}
b \mapsto b_\star\equiv \frac{b}{\sqrt{1 + {b^2}/{b_{\text{max}}^2}}}\,.
\end{equation}
Although this prescription guarantees that the variable $b_\star$ saturates at $b_{\text{max}}$ at large values of $b$, 
then avoiding the Landau singularity, 
it also introduces a nonphysical dependence on $b_{\text{max}}$ in the form of spurious power corrections that must be compensated for
(we will comment below).

The finite component of the distribution in Eq.\,\eqref{e:dsdz} is a process-dependent function, which is obtained from the fixed-order result by subtracting the perturbative expansion of the resummed contribution, truncated at the same perturbative order:
\begin{equation}
\frac{1}{\sigma_{tot}} \frac{d\Sigma_{(fin.)}}{d z}
= \frac{1}{\sigma_{tot}} \frac{d\Sigma_{(f.o.)}}{d z}
- \left.
\frac{1}{\sigma_{tot}} \frac{d\Sigma_{(res.)}}{d z}
\right|_{\text{f.o.}} \, .
\end{equation}
For $z>0$, the finite remainder admits a
standard perturbative expansion
\begin{equation}
\frac{1}{\sigma_{tot}} \frac{d\Sigma_{(fin.)}}{d z}
= \frac{\alpha_S}{\pi}\, \widetilde{\mathcal{A}}_{(fin.)}(z)
+ \Big( \frac{\alpha_S}{\pi} \Big)^2 \widetilde{\mathcal{B}}_{(fin.)}(z)
+ \Big( \frac{\alpha_S}{\pi} \Big)^3 \widetilde{\mathcal{C}}_{(fin.)}(z)
+ \mathcal{O}(\alpha_S^4)\, ,
\end{equation}
where the coefficients 
of the powers of $\alpha_S/\pi$
have been explicitly given
in~\cite{Aglietti:2024xwv}
up to the next-to-next-to-leading order (NNLO),
i.e.\ $\mathcal{O}(\alpha_S^3)$.

\subsection{Non-perturbative model}

We incorporate non-perturbative effects 
in our theoretical description of the EEC,
by multiplying the Sudakov form factor $S(b,Q)$ 
by the following NP form factor 
\begin{eqnarray}
\label{e:SNP}
    S_{NP}\left(b,\frac{Q}{Q_0}\right) &=& 
    \exp\left[- f(b) - g_K(b) \,
    \ln\left({\frac{Q^2}{Q_0^2}}\right)\right]\,.
\end{eqnarray}
The function $f(b)$ parameterizes the non-perturbative effects at very large values of $b$ ($b \gtrsim 1~\text{GeV}^{-1}$) 
at a reference energy scale $Q_0$, while the function $g_K(b)$ 
encodes the NP evolution from $Q_0$ to $Q$ and is thus the NP component of the so-called Collins--Soper evolution
kernel~\cite{Collins:1981va,Collins:1985xx,Collins:1984kg,Collins:2011zzd}. By requiring NP effects to vanish in the perturbative region (i.e.\ at small $b$),
we impose $S_{NP}(b=0,Q/Q_0)=1$ which implies $f(0)=g_K(0)=0$. In this work, we set $Q_0=1$~GeV
and use the following parameterization
\begin{align}
\label{e:NP1}
    f(b) &= f_1\, b +f_2\,b^2\, , \\
    g_K(b) &=g_0 \left\{ 1 - \exp \Big{[} - \frac{C_F \, \alpha_S \, (b_0^2/b_\star^2)}{\pi g_0}
    \frac{b^2}{b^2_{\text{max}}}
    \Big{]} \right\}\,;    
    \label{e:NP2}
\end{align}
where $g_0$, $f_1$ and $f_2$ are NP parameters to be extracted from the experimental data. 
The functional form in Eq.\,\eqref{e:NP1}, with a linear and a quadratic term in $b$, is the same dictated by the Dokshitzer-Marchesini-Webber (DMW) dispersive  model~\cite{Dokshitzer:1999sh,Dokshitzer:1995qm} and successfully used in 
Ref.\,\cite{Aglietti:2024xwv} to describe the data at the $Z$-boson resonance.  
The functional form of $g_K(b)$ in Eq.\,\eqref{e:NP2} has been proposed in Ref.\,\cite{Collins:2014jpa} 
and has been successfully used in global analysis of the Drell--Yan transverse-momentum distribution at various
center-of-mass energies~\cite{Camarda:2025lbt} but also in a precise determination of $\alpha_S$ in Ref.\,\cite{Camarda:2022qdg,ATLAS:2023lhg}. 
It has been chosen in order to cancel the leading $\mathcal{O}(b^2/b_{\text{max}}^2)$ dependence of the perturbative form  factor $S(Q,b)$ from $b_{\text{max}}$, leaving a residual $\mathcal{O}(b^4/b_{\text{max}}^4)$ dependence. In this work we use $b_{max}= b_0=2 \exp\{-\gamma_E\}\simeq 1.123\,\text{GeV}^{-1}$.

\subsection{Heavy-quark mass effects}

In Ref.\,\cite{Aglietti:2024zhg}, heavy-quark contributions to the Sudakov form factor of the EEC distribution were computed. In this work, we investigate the impact of these effects on our analysis. In particular, by including finite-mass effects only for the bottom quark, we rewrite the form factor $S(b,Q)$ entering Eq.\,\eqref{e:resum} as~\cite{Aglietti:2024zhg}:
\begin{eqnarray}
S(b,Q) &\mapsto& S(b,Q) \, \theta(b_{\text{cr}} -b) + 
 \Bigg{[} \frac{c_b(Q^2)}{\sum_f c_f(Q^2)} S(b_{\text{cr}},Q) S_m(b,Q)\vert_{n_f=4}
\, \notag\\
&+&
\left( 1- \frac{c_b(Q^2)}{\sum_f c_f(Q^2)}\right) S(b,Q)\Big\vert_{n_f=4} \Bigg{]}
\,\theta(b-b_{\text{cr}}),
\label{e:resum_mass}
\end{eqnarray} 
where $b_{\text{cr}} \equiv b_0 /m_b$ is the critical length below which the effects of the bottom-quark mass
$m_b$ are neglected and $S_m(b,Q)$ is the massive Sudakov form factor, whose explicit expression can be found in \cite{Aglietti:2024zhg}. 
The subscript $n_f=4$ denotes the terms in which the number of active quark flavors $n_f$ has been reduced from five to four,
in order to remove the contribution from a massive (real or virtual) heavy quark pair. 
The sum in Eq.\,\eqref{e:resum_mass} runs over all active quark flavors,
$f=u,d,c,s,b$, and $c_f$ is the electroweak
coefficient related to quark flavor $f$, 
given, in the massless case, by~\cite{Ellis:1996mzs}
\begin{equation}
\label{e:EW_charges}
c_f(Q^2) = Q_f^2 - 2 V_e Q_f V_f \, \chi_1(Q^2) + (V_e^2 + A_e^2)\, (V_f^2 + A_f^2)\, \chi_2(Q^2),
\qquad f = u,d,s,c \; .
\end{equation}
In the bottom-quark case, by including leading mass effects:
\begin{eqnarray}
c_b(Q^2) \!&=&\! 
\bigg\{
\Big[
Q_b^2 - 2 V_e Q_b V_b \, \chi_1(Q^2)
+ \left(V_e^2+A_e^2\right) V_b^2 \, \chi_2(Q^2)
\Big] 
\beta \frac{3-\beta^2}{2}
\left[
1 + \left( c_1 - 1 \right) \frac{\alpha_S}{\pi}
\right]
\nonumber\\
&+&
\left(V_e^2 + A_e^2\right) A_b^2 \, \chi_2(Q^2)
\beta^3
\left[
1 + \left( d_1 - 1 \right) \frac{\alpha_S}{\pi}
\right]
\bigg\}\,\theta(Q-2m_b).
\end{eqnarray}
$\theta(x)\equiv 1$ for $x>0$
and zero otherwise is the Heaviside step function
and $\beta$ is the (ordinary) velocity of the bottom quark,
\begin{equation}
\beta = \sqrt{1-\frac{4m_b^2}{Q^2}},
\end{equation}
with $m_b$ the on-shell bottom-quark mass.
The coefficients $c_1$ and $d_1$ are 
the $\mathcal{O}(\alpha_S)$ 
polar-vector and axial-vector corrections to the 
total cross section of $e^+e^-\to b \bar{b}$
in the massive case, respectively:
\begin{eqnarray}
c_1 &=& 1 + 12\, \epsilon + \mathcal{O}(\epsilon^2);
\nonumber\\
d_1 &=& 1 - 22\, \epsilon + \mathcal{O}(\epsilon^2);
\end{eqnarray}
where:
\begin{equation}
\epsilon \equiv \frac{\overline{m}_b^2(Q^2)}{Q^2} \ll 1.
\end{equation}
$\overline{m}_b(Q^2)$ is the  $\overline{\text{MS}}$
(running) bottom-quark mass at the renormalization scale
$\mu=Q$. 
The functions $\chi_i$, $i=1,2$, 
involving the $Z^0$-exchange diagram,
explicitly read:
\begin{align}
\label{e:EW_chi_functions}
\chi_1(Q^2) &= \frac{1}{4 \sin^2\theta_W \cos^2\theta_W } \frac{Q^2 ( Q^2 -  m_Z^2 )}{ (Q^2 - m_Z^2)^2 + m_Z^2 \Gamma_Z^2} \; ,\\
\chi_2(Q^2) &= \frac{1}{16 \sin^4\theta_W\cos^4\theta_W} \frac{Q^4}{ (Q^2 - m_Z^2)^2 + m_Z^2 \Gamma_Z^2} \; ,
\end{align}
where the constants $Q_f$, $V_f$ and $A_f$ represent the electric, vector, and axial charges of flavor $f$ respectively; $V_e$ and $A_e$ are the vector and axial charges of the electron; $\theta_W$ is the weak mixing angle; $\Gamma_Z$ and $m_Z$ are the width and mass of the $Z$ boson.
We use the following numerical values: 
$\Gamma_Z=2.4950\,\text{GeV}$, $m_Z=91.1876\,\text{GeV}$, $\sin^2\theta_W=0.23121$, $m_b=4.2\,\text{GeV}$~\cite{ParticleDataGroup:2024cfk}.

\section{Data and fitting procedure}
\label{s:analysis}

The experimental data set used in this analysis
includes all available measurements 
of the EEC function, including the recent reanalysis of the ALEPH Coll.\,\cite{Electron-PositronAlliance:2025wzh,Electron-PositronAlliance:2025fhk} at $\sqrt{s} = 91.2$ GeV and the analysis of the AMY Coll.\,\cite{AMY:1997waz} at $\sqrt{s} = 58.0$ GeV. 
Notably, these specific datasets are included here for the first time in a global EEC analysis, providing unique constraints on the energy evolution of the distribution. 
We find that our theoretical framework provides an excellent description of these additional datasets, with a $\chi^{2}$ per degree of freedom of the order of unity.
%\LRcom{
\\
Let us remark that the ALEPH analysis includes only charged final-state 
particles and therefore differs from the fully inclusive EEC observable. 
In Sec.~\ref{s:results} we discuss the origin of this difference and show 
that it is numerically negligible. As a further check, we perform a fit 
without the ALEPH dataset and show that the extracted parameters do not 
significantly change. Finally, we show that our theoretical framework 
provides an excellent description of the ALEPH data, with a $\chi^2$ per 
degree of freedom of order unity, both when the dataset is directly 
included in the fit and when a postdiction is performed to assess the 
level of agreement with the other measurements.
%}
%, these data are not included in the fit. Instead, after determining the best-fit parameters, we provide postdictions to assess the level of agreement with the measurements.
%are included here for the first time in a global EEC analysis. 

Following Ref.~\cite{Aglietti:2024xwv}, in our numerical analysis we perform the fit by consistently including data points from both the peak and the intermediate regions. This strategy ensures a reliable extraction of the non-perturbative parameters, as well as a robust determination of the strong coupling constant. We therefore restrict the analysis to the region of intermediate and large values of the $\chi$ kinematic variable\footnote{We have verified that our results are stable under small variations of this range.}:
\begin{equation}
1.8 \leq \chi \leq \pi \, . 
\end{equation}
In total, 691 data points are included in the fit, which decrease to 610 if the ALEPH dataset is excluded. The relevant information for each dataset is summarized in Table~\ref{t:data}.

\begin{table}[th]
\footnotesize
\begin{center}
\renewcommand{\tabcolsep}{0.3pc}
\renewcommand{\arraystretch}{1.5}
\begin{tabular}{|c|c|c|c|c|c|}
  \hline
  Experiment & $N_{\rm dat}$   &  $\sqrt{s}$ [GeV] & Ref. \\
  \hline
  \hline
  OPAL (1992) & 43 & 91.3  & \cite{OPAL:1991uui} \\
    \hline
  OPAL (1993) & 77 & 91.2  & \cite{OPAL:1993pnw} \\
  \hline
  DELPHI (1992) & 21 & 91.2  & \cite{DELPHI:1992qrr} \\
  \hline
  DELPHI (1993) & 21 & 91.2  & \cite{DELPHI:1993nlw} \\
  \hline
  L3 & 15 & 91.2  & \cite{L3:1992btq} \\
  \hline
  SLD & 21 & 91.2  & \cite{SLD:1994idb} \\
    \hline
    ALEPH & 81 & 91.2  & \cite{Electron-PositronAlliance:2025wzh,Electron-PositronAlliance:2025fhk}\\
  \hline
    \makecell{TOPAZ} & \makecell{21\\21} & \makecell{59.5 \\ 53.3} & \cite{TOPAZ:1989yod} \\
\hline
    \makecell{AMY} & \makecell{43} & \makecell{58.0} & \cite{AMY:1997waz} \\
\hline
    \makecell{TASSO} & \makecell{21\\21\\21\\21} & \makecell{43.5\\34.8\\22.0\\14.0} & \cite{TASSO:1987mcs} \\
  \hline
    PLUTO &  13 & 34.6 & \cite{PLUTO:1985yzc} \\
    \hline
    \makecell{PLUTO} & \makecell{8\\8\\8\\8\\8\\8\\8\\6} & \makecell{30.8\\27.6\\22\\17\\13\\12\\9.4\\7.7} & \cite{PLUTO:1981gcc} \\
  \hline
    \makecell{JADE} & \makecell{21\\21\\21} & \makecell{34.0\\22.0\\14.0} & \cite{JADE:1984taa} \\
  \hline
    \makecell{CELLO} &  \makecell{21\\21} &\makecell{34.0\\22.0} & \cite{CELLO:1982rca} \\
  \hline
    \makecell{MARKII Run I} &  21 & 29.0 & \cite{Wood:1987uf} \\
  \hline
    \makecell{MARKII Run II} &  21 & 29.0 & \cite{Wood:1987uf} \\
  \hline
    MAC &  21 & 29.0 & \cite{Fernandez:1984db} \\
  \hline
  \hline
%  Total & 610  & & \\
%      \hline
%      \hline
%  \hline
    Total & 691  & & \\
      \hline
\end{tabular}
\caption{Experimental data sets included in this analysis. Each row contains the name of the collaboration, the number of data points ($N_{\rm dat}$) included, the center-of-mass energy $\sqrt{s}$ and the reference.}
\label{t:data}
\end{center}
\end{table}
As pointed out in Refs.~\cite{Kardos:2018kqj,Cuerpo:2025zde}, a key aspect of a global analysis of EEC data is the treatment of experimental uncertainties. Here we follow the approach adopted in Ref.~\cite{Kardos:2018kqj}. In particular, for datasets where statistical and systematic uncertainties are provided separately, we treat them consistently. For datasets where only partial information on systematic uncertainties is available, we estimate them following the procedure of Ref.~\cite{Kardos:2018kqj}. For all remaining datasets, the published combined uncertainties are treated as purely statistical.
\\
Another important aspect discussed in Ref.~\cite{Kardos:2018kqj} is the presence of correlations between different bins within the same experiment. Since correlation matrices are not available for all datasets, we assess their impact by considering two scenarios. In the first, no correlations are included, reflecting the information directly available. In the second, following Ref.~\cite{Kardos:2018kqj}, we assume a correlation coefficient $\rho = 0.5$ between adjacent bins of the same experiment, for both statistical and systematic uncertainties.  The associated uncertainty is then estimated from the spread of the fitted parameters obtained in the two scenarios.
We perform a fit to determine the optimal values of $\alpha_S(m_Z^2)$ and of the non-perturbative parameters $f_1$, $f_2$, and $g_0$ entering Eq.~\eqref{e:SNP}. The best-fit parameters are obtained by minimizing the following $\chi^2$ function:
\begin{equation}
\chi^2 = \sum_{d \in \text{datasets}} \chi^2_d 
= \sum_{d \in \text{datasets}} 
\left( \vec{D}_d - \vec{P}_d \right)^T 
V_d^{-1} 
\left( \vec{D}_d - \vec{P}_d \right) \, ,
\end{equation}
where the index $d$ runs over all available datasets. The vector $\vec{D}_d$ denotes the data points of the $d$-th dataset, $V_d$ is the corresponding covariance matrix, and $\vec{P}_d$ represents the theoretical predictions for that dataset.

In order to obtain a reliable estimate of the uncertainties on 
$\alpha_S(m_Z^2)$ and on the non-perturbative parameters arising from experimental uncertainties, we employ the \textit{bootstrap} method. This method, originally introduced for collinear PDF fits~\cite{Ball:2008by,Ball:2012wy}, consists in generating replicas of the dataset by adding Gaussian fluctuations according to the experimental uncertainties and fitting each replica independently.  
We generate 1000 replicas, obtaining 1000 sets of best-fit parameters. By construction, their mean values coincide with the central values obtained from the fit to the original dataset, while their standard deviations provide the corresponding uncertainties at the 68\% confidence level. Further details can be found in Refs.~\cite{Ball:2008by,Ball:2012wy}. 
To estimate the size of missing higher-order contributions and the associated perturbative uncertainties, we consider conventional variations of the auxiliary scales $\mu_R$ and $\mu_Q$. In particular, we perform fits by varying $\mu_R$ and $\mu_Q$ independently within the range
\begin{equation}
\label{e:scale}
\frac{Q}{2} \leq \{\mu_R,\mu_Q\} \leq 2Q \, ,
\end{equation}
subject to the constraint
\begin{equation}
\frac{1}{2} \leq \frac{\mu_R}{\mu_Q} \leq 2 \, .
\end{equation}
At the end of the analysis, the fitted parameters are assigned three different sources of uncertainty: experimental, correlation-related, and perturbative. The total uncertainty for $\alpha_S$ (and analogously for the non-perturbative parameters) is expressed as
\begin{equation}
\label{e:scale2}
\delta\alpha_S = \sqrt{(\delta \alpha_S^{\mathrm{exp}})^2 + (\delta \alpha_S^{\mathrm{corr}})^2 + (\delta \alpha_S^{\mathrm{th}})^2 } \, ,
\end{equation}
where %$\overline{\alpha_S}$ denotes the central value obtained from the bootstrap method, 
$\delta \alpha_S^{\mathrm{exp}}$ represents the uncertainty due to experimental fluctuations, $\delta \alpha_S^{\mathrm{corr}}$ accounts for the modeling of bin-to-bin correlations, and $\delta \alpha_S^{\mathrm{th}}$ denotes the perturbative uncertainty estimated from the envelope of the fits obtained under scale variations.
\clearpage
\section{Results}
\label{s:results}
In this section, we present the results obtained by applying the formalism described in the previous sections at N$^3$LL+NNLO accuracy. 

%\LRcom{
In general, we find an excellent description of the data, as indicated by the value $\chi^2 / N_{\text{d.o.f.}} = 1.2$, obtained from the fit to the unfluctuated (central) dataset without including any correlations (i.e.\ $\rho = 0$).
We have also investigated the impact of bin-to-bin correlations by performing an alternative fit that assumes a correlation coefficient $\rho = 0.5$ between adjacent bins within the same dataset, as described in Sec.~\ref{s:analysis}. In this case, we obtain $\chi^2 / N_{\text{d.o.f.}} = 1.4$, which is compatible with the uncorrelated result. This indicates that our conclusions are stable under reasonable assumptions about the correlation structure. The variations in the fitted parameters are included in the total uncertainty, as described in the previous section, and are reported in Table~\ref{t:par}.
%}

\begin{figure}
\centering
\includegraphics[width=0.49\textwidth]{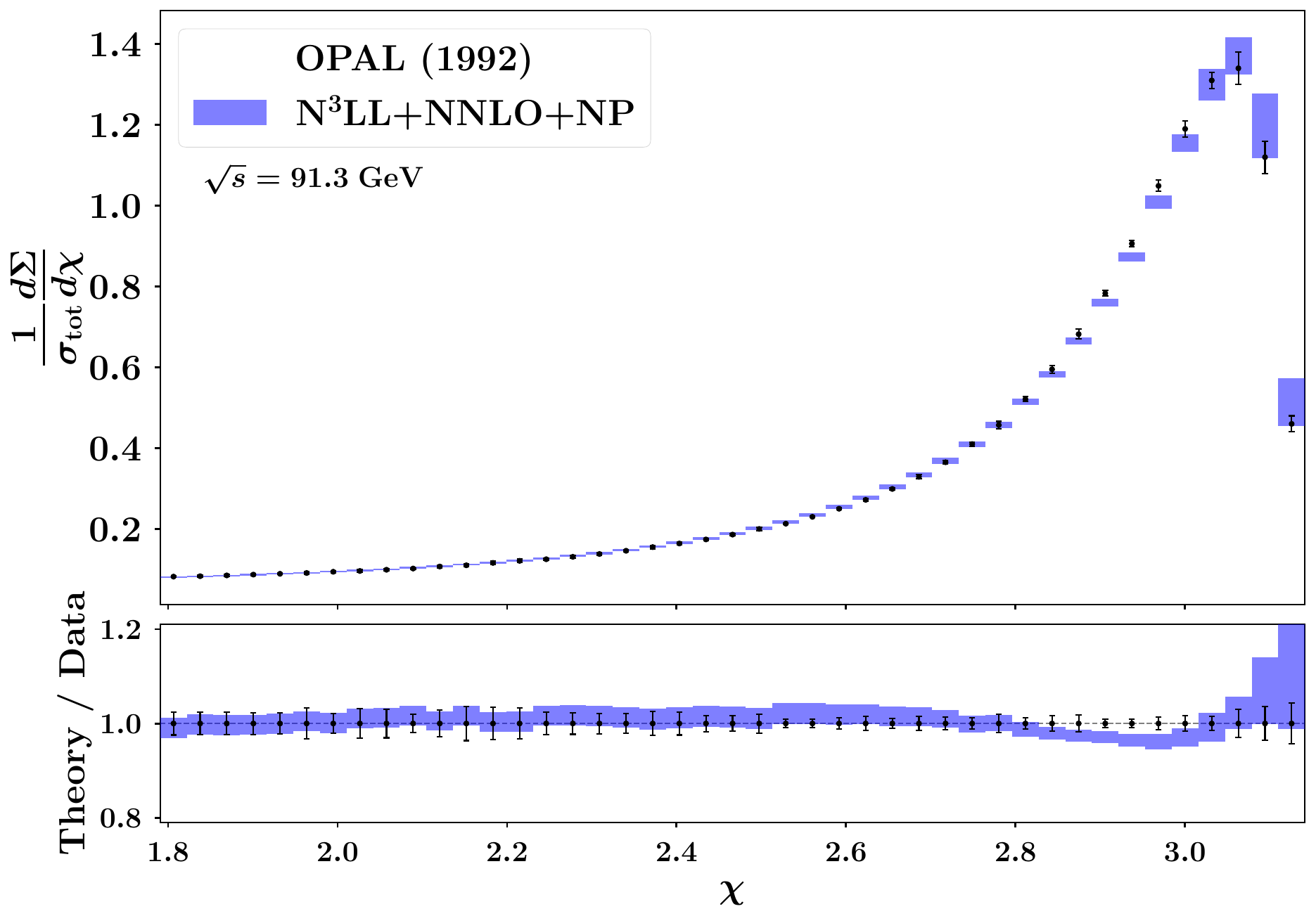}
\includegraphics[width=0.49\textwidth]{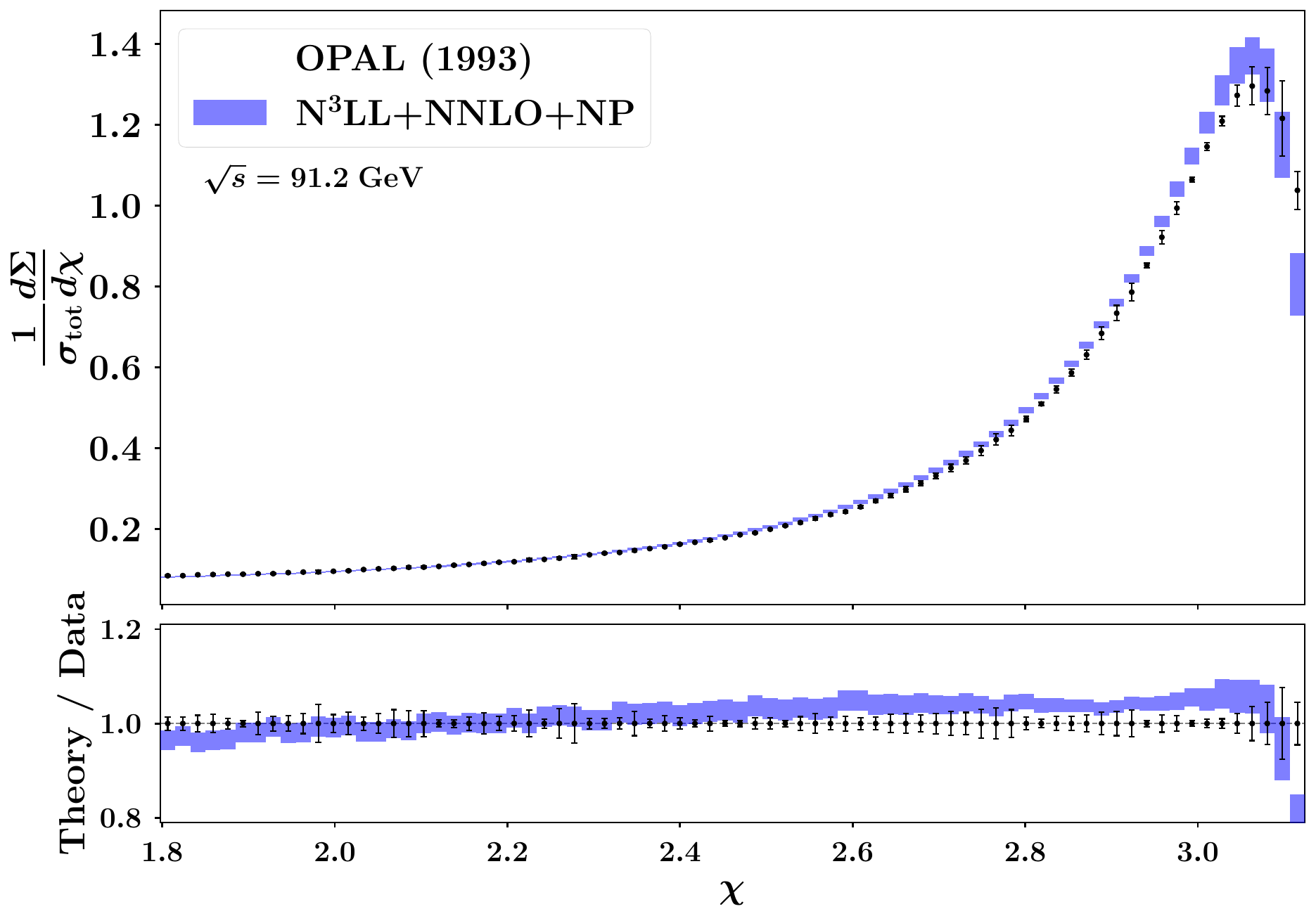}
\includegraphics[width=0.48\textwidth]{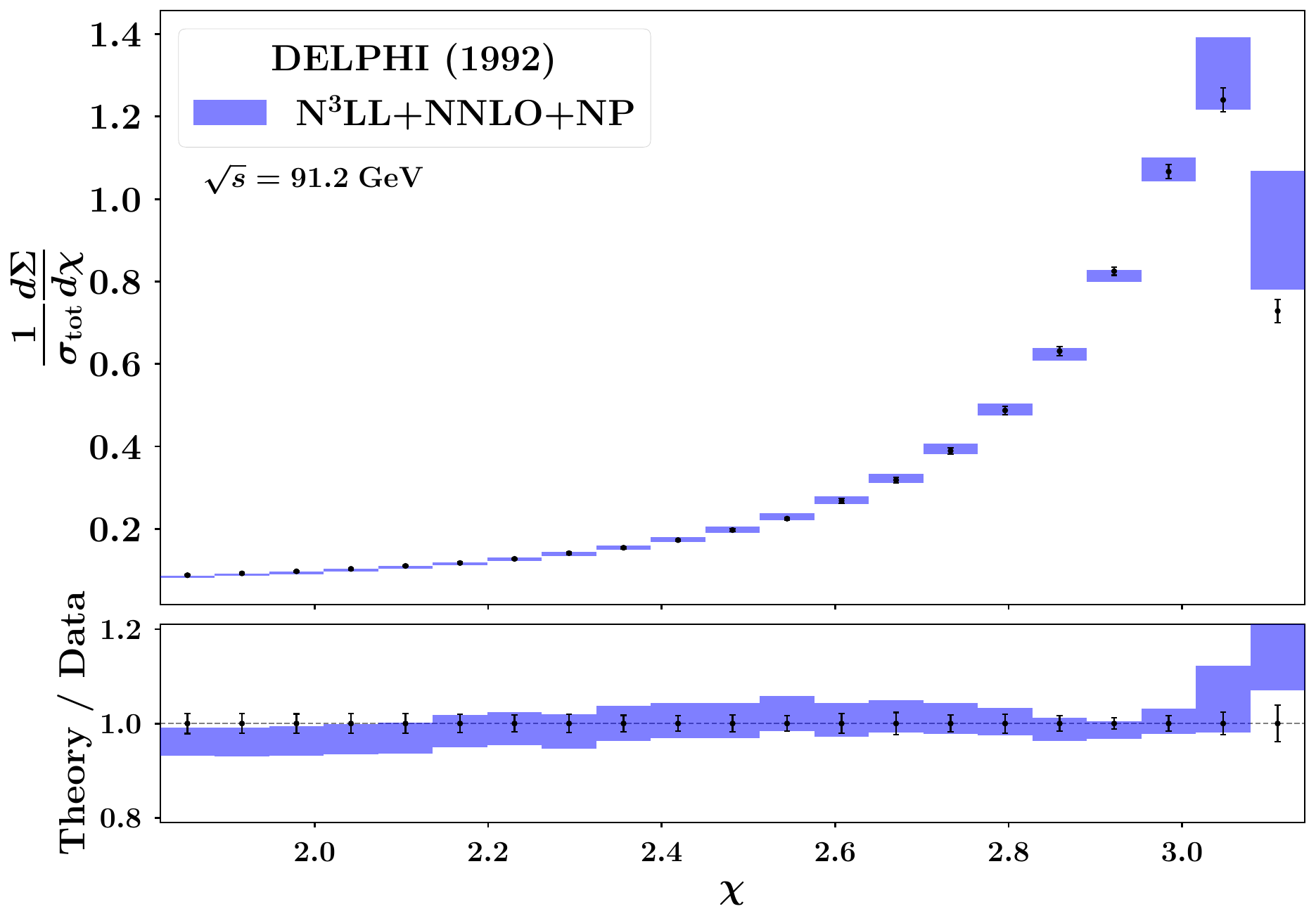}
\includegraphics[width=0.48\textwidth]{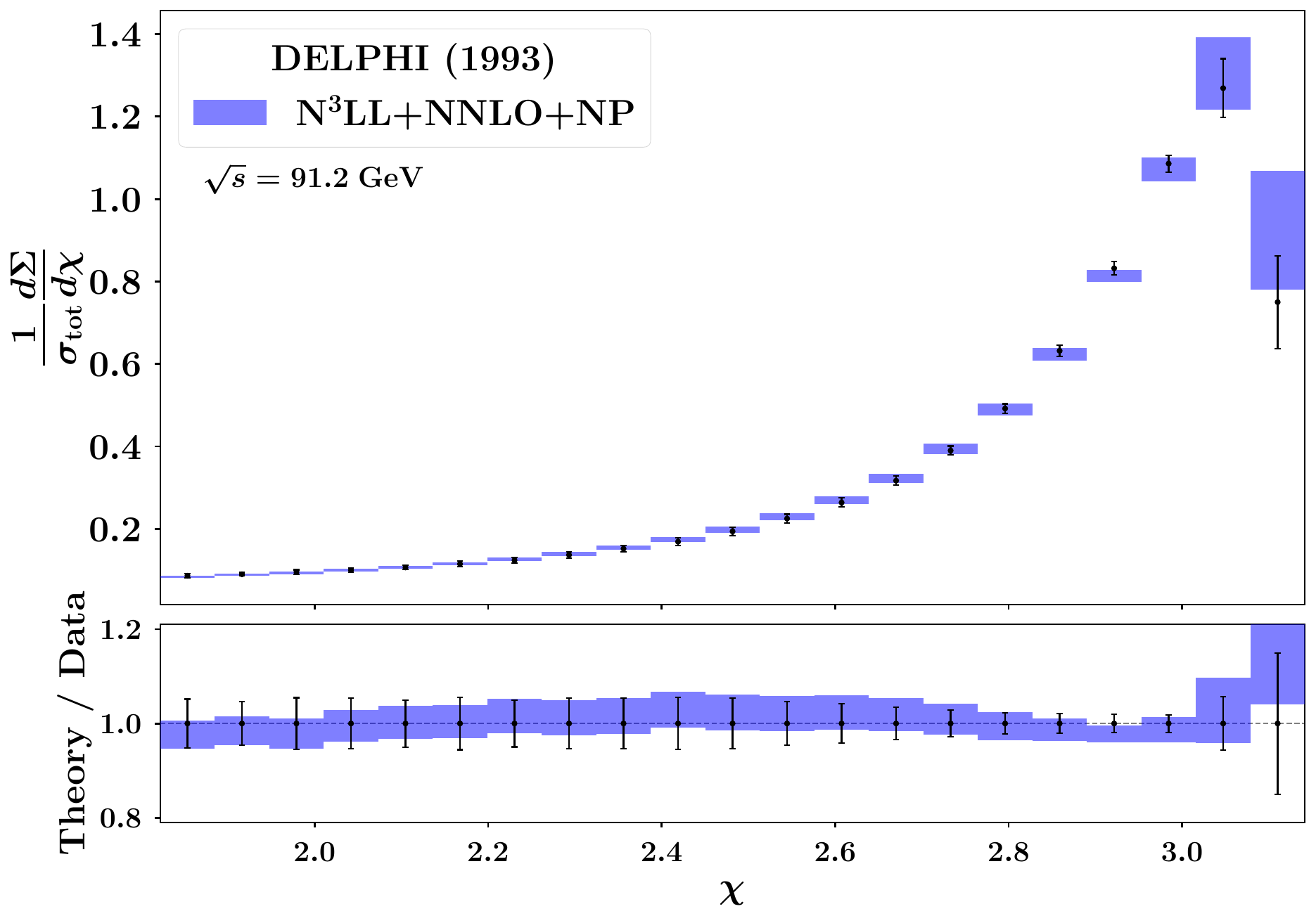}
\includegraphics[width=0.48\textwidth]{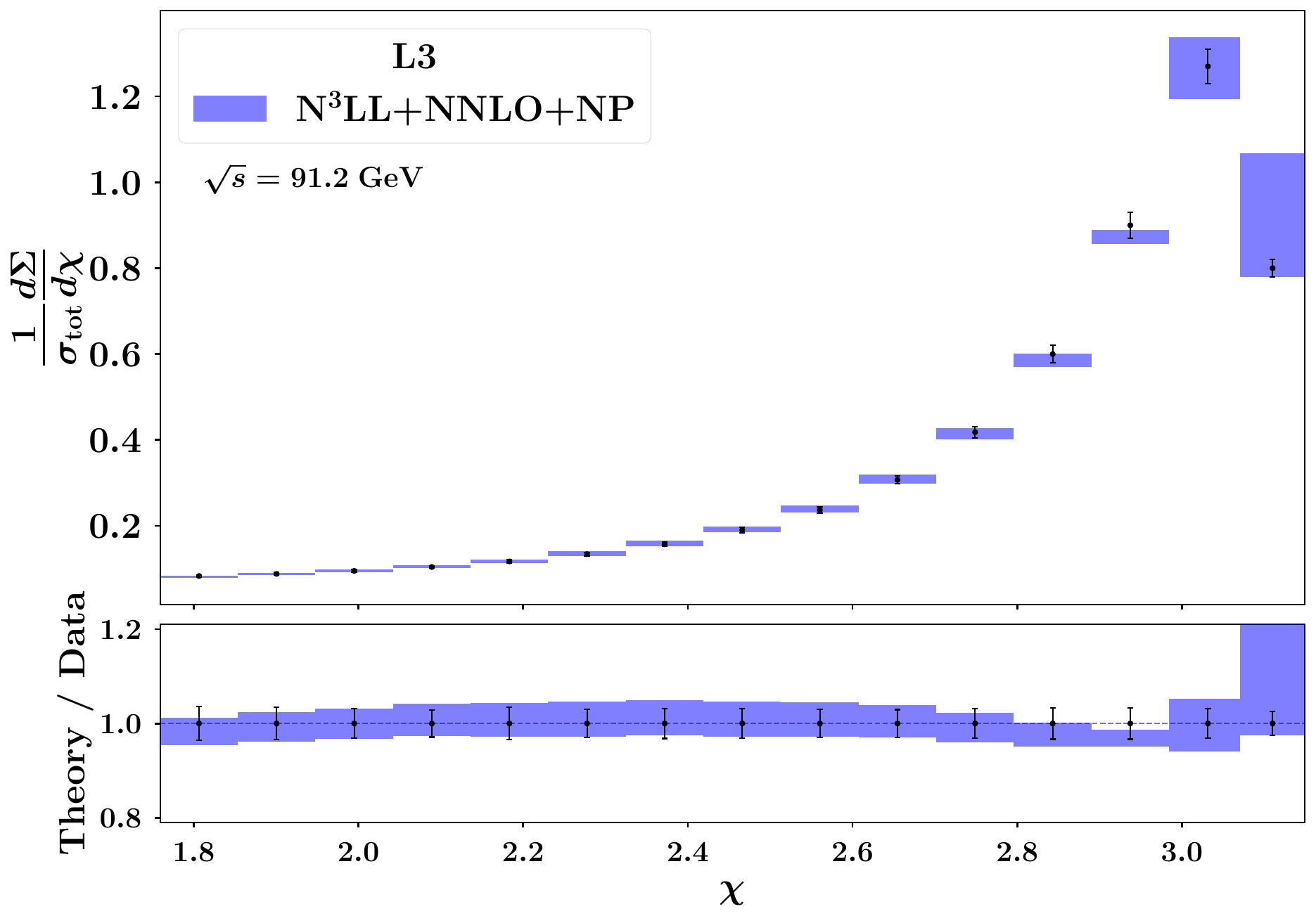}
\includegraphics[width=0.48\textwidth]{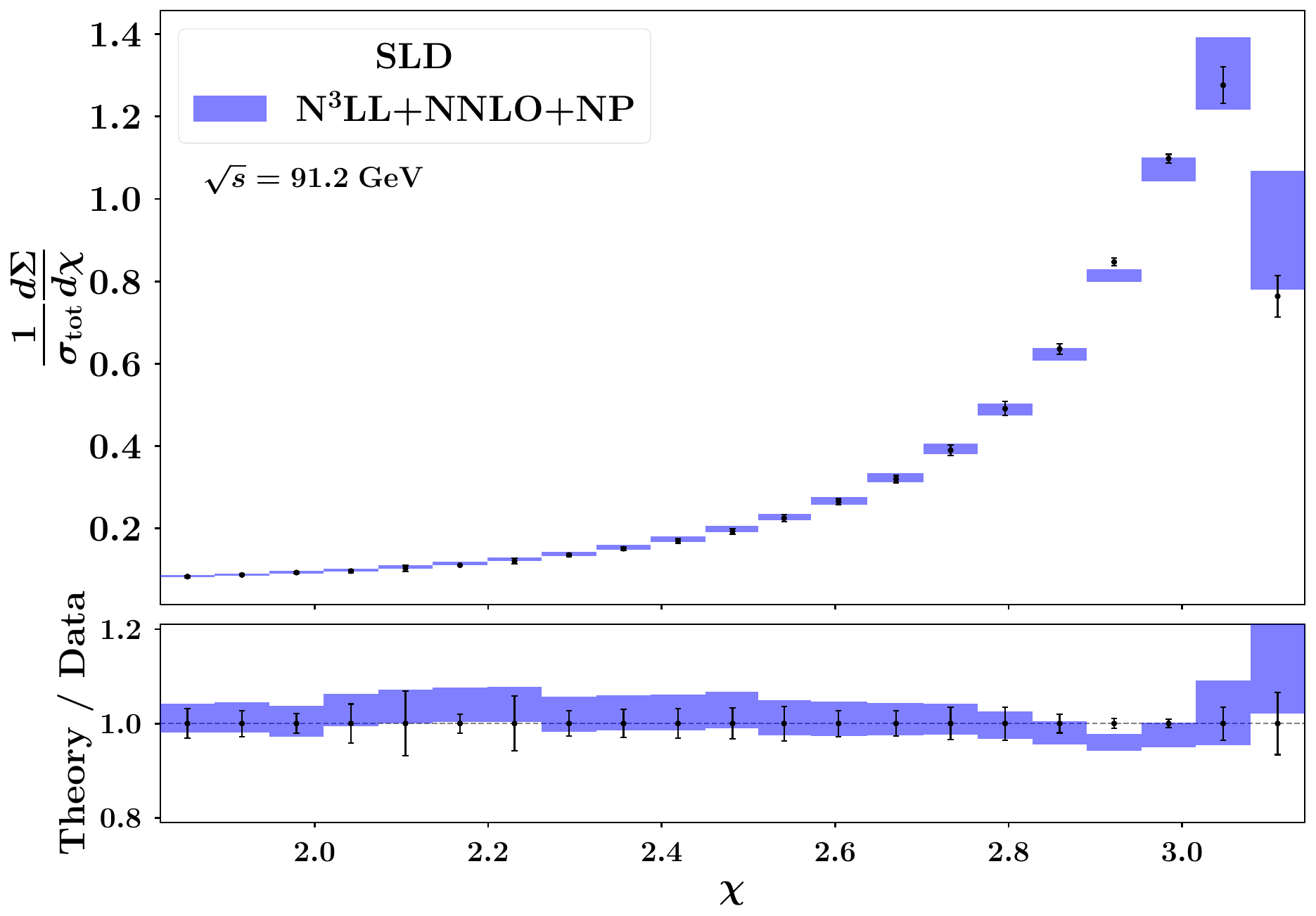}
\caption{Comparison between experimental data and theoretical predictions (blue bands) for the EEC distribution 
at the $Z$-boson resonance from OPAL Coll.\ (top left and top right),
DELPHI Coll.\ (middle left and right), L3 Coll.\ (bottom left) and SLD Coll.\ (bottom right). The uncertainty bands represent the theoretical uncertainties estimated through scale variations. The lower panels show the ratio of the theoretical results to the experimental data.}
\label{f:SLD-DELPHI-L3}
\end{figure}

In Fig.~\ref{f:SLD-DELPHI-L3} we compare the theoretical predictions obtained from the global fit with experimental data from LEP (CERN) and SLC (SLAC)~\cite{OPAL:1991uui,OPAL:1993pnw,DELPHI:1992qrr,L3:1992btq,SLD:1994idb}, corresponding to measurements at the $Z$-boson mass. The blue uncertainty bands represent the estimated theoretical uncertainties, evaluated through standard scale variations (see Eqs.~\eqref{e:scale} and \eqref{e:scale2}).
The experimental data are well described, within uncertainties, over the entire selected angular range. This provides clear evidence that the formalism described in Sec.~\ref{s:formalism}, supplemented by an analytic non-perturbative model, is able to accurately describe the data. Moreover, the uncertainty bands are comparable to those obtained from a fit using only $Z$-pole data~\cite{Aglietti:2024xwv}. This indicates the absence of significant tensions between measurements at the $Z^0$ peak and at lower energies.

\begin{figure}
\centering
\includegraphics[width=0.49\textwidth]{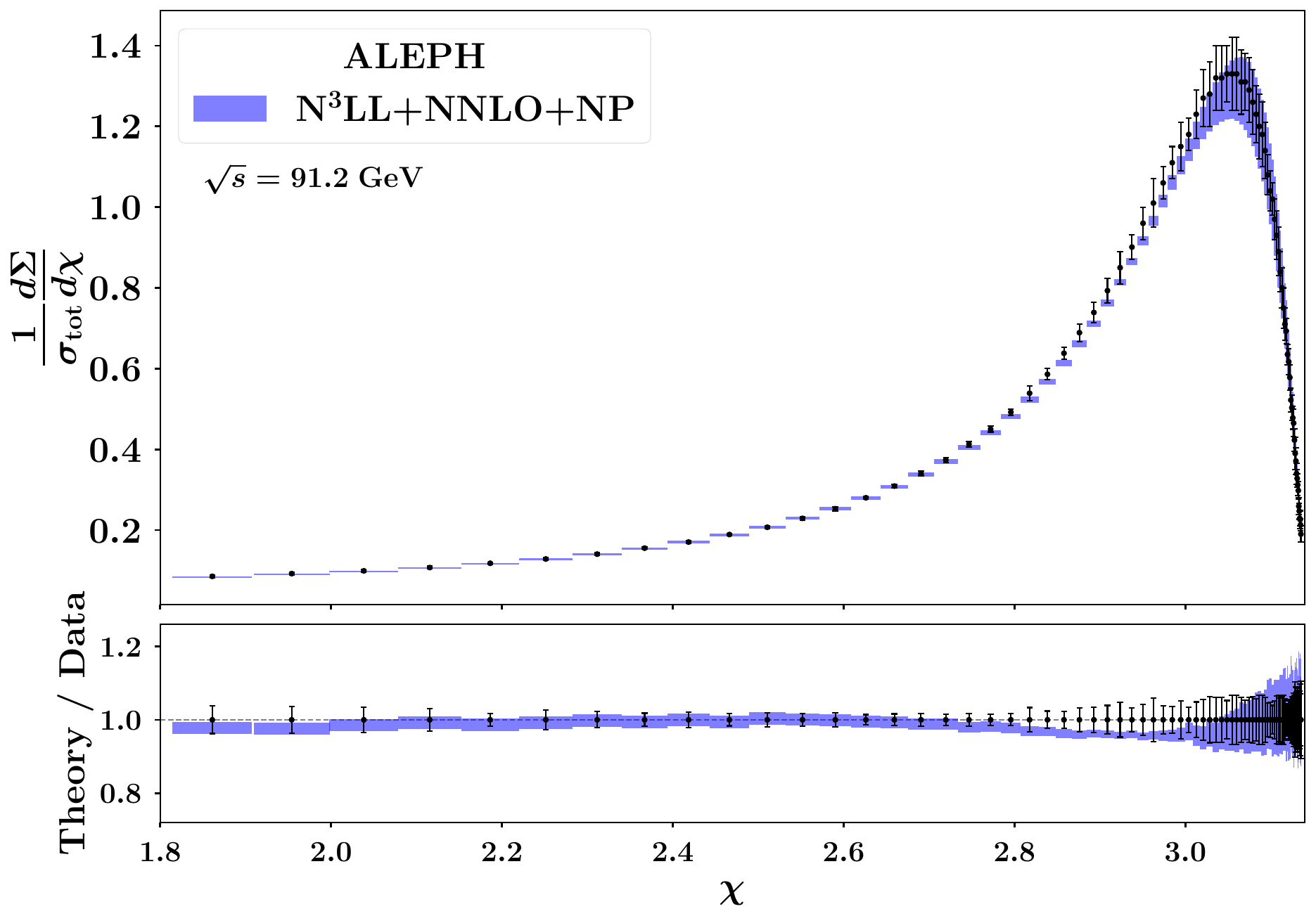}
\includegraphics[width=.49\textwidth]{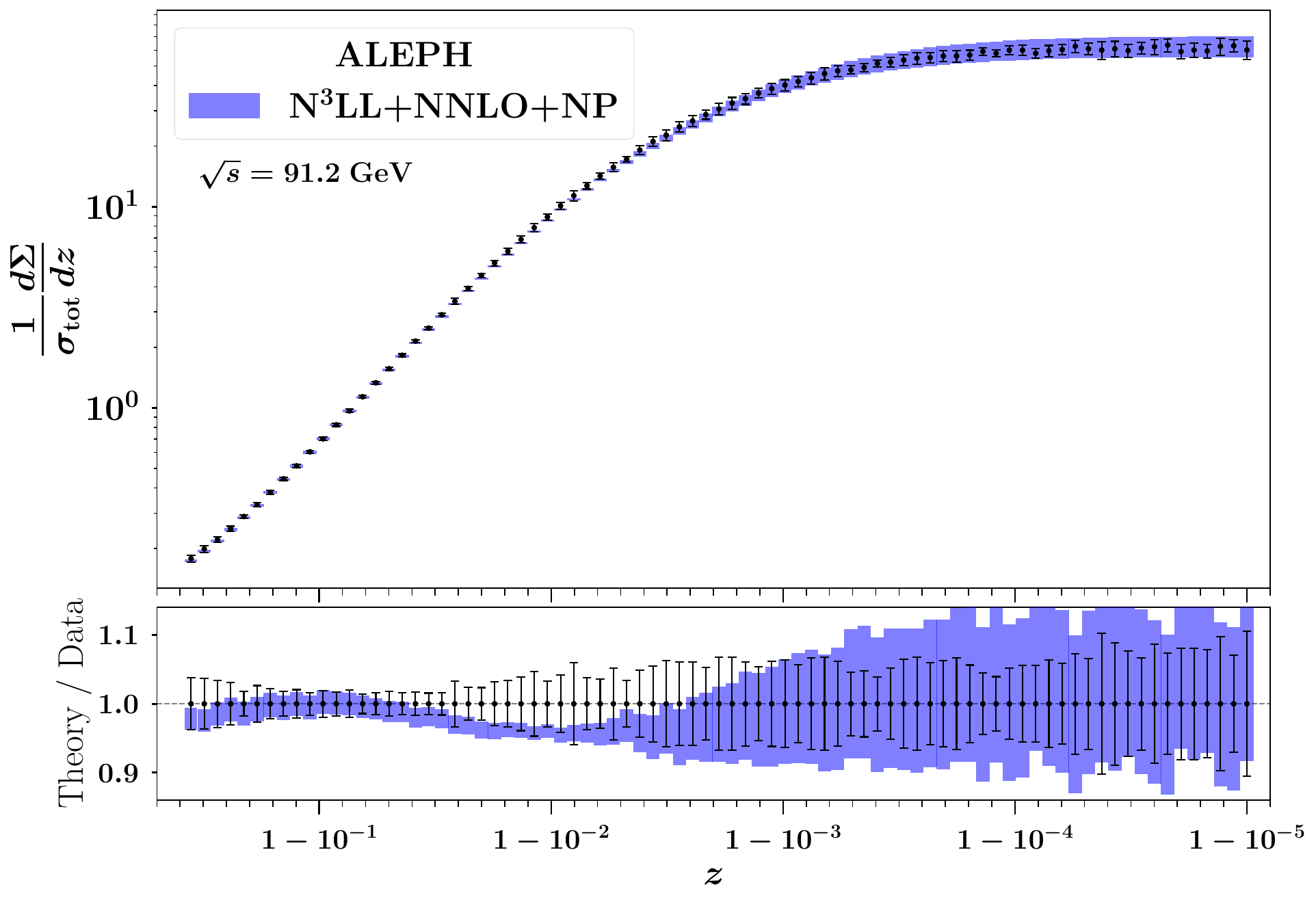}
\caption{Comparison between the recent reanalysis of archived experimental data from ALEPH Coll.\ at 
$\sqrt{s}=91.2\,\text{GeV}$ and our theoretical predictions (blue bands).
The EEC distribution is shown as a function of $\chi$ (left panel) and as a function of $z$ on logarithmic scales (right panel), the latter emphasizing the back-to-back region where resummation and non-perturbative effects are most prominent. 
The uncertainty bands represent the theoretical uncertainties estimated through scale variations. The lower panels show the ratio of the theoretical results to the experimental data. 
The superscript {\itshape ch.} 
indicates that the ALEPH analysis is based on charged final-state particles only, rather 
than on the fully inclusive hadronic final state.
}
\label{f:ALEPH}
\end{figure}

One of the main novelties of this work is the inclusion of the recent reanalysis of archived data from the ALEPH Coll.\ at the 
$Z$-boson resonance \cite{Electron-PositronAlliance:2025wzh,Electron-PositronAlliance:2025fhk}\,\footnote{It is worth emphasizing that the experimental data from the ALEPH collaboration have been accurately self-extracted from the original publications, as the corresponding numerical tables are not available in public databases.}. 
In Fig.\,\ref{f:ALEPH}, we compare our theoretical predictions with these recent high-precision measurements. 
The EEC distribution is shown both as a function of $\chi$ and as a function of $z$ on logarithmic scales.
By going from the variable $\chi$ to 
the variable $z$, the back-to-back region, which is our primary concern,
is strongly compressed, and the peak disappears.
Our N$^3$LL+NNLO predictions, supplemented by the analytic dispersive model, achieve a very good description of this novel dataset across the entire available spectrum. This remarkable agreement, also in the case of smaller experimental uncertainties, further confirms the robustness of our theoretical framework.

The ALEPH dataset differs from the other measurements included in this 
analysis in that it is based on charged final-state particles only, rather 
than on the fully inclusive hadronic final state. Strictly speaking, the 
theoretical framework described in Sec.~\ref{s:formalism} computes the 
EEC summed over all final-state hadrons. However, as we discuss below, the 
difference between the charged-particle EEC and the fully inclusive EEC 
is numerically negligible in the kinematic region relevant for our fit.
\\
The charged and the inclusive EEC are related by
\begin{equation}
\frac{1}{\sigma^{ch}_{\rm tot}}\frac{d\Sigma^{\rm ch}}{d\cos\chi}
\;=\;
\frac{1}{\sigma_{\rm tot}}\frac{d\Sigma}{d\cos\chi} \cdot 
\bigl[1 + \varepsilon(\chi)\bigr] \,,
\label{eq:charged_vs_all}
\end{equation}
where $\sigma_{\rm tot}^{\rm ch}$ denotes the charged-weighted total cross section,
defined as
\begin{equation}
    \sigma_{\rm tot}^{\rm ch} = \int d\sigma \Bigg{(}\sum_{i\in \text{ch}} \frac{E_i}{Q}\Bigg{)}^2
\end{equation}
and $\varepsilon(\chi)$ 
parameterizes the residual $\chi$-dependent corrections.

The subleading correction $\varepsilon(\chi)$ receives contributions from
two distinct non perturbative sources. The first is associated with
event-by-event fluctuations of the charged energy fraction,
$r \equiv \sum_{i\in {\rm ch}} E_i/Q$, around its mean value. The second arises from the
$\chi$-dependent isospin composition of the hadronic final state, which
is governed by hadronization dynamics. The fluctuation contribution is
suppressed by the large final-state multiplicity at LEP energies, where
$\langle n_{\rm ch}\rangle \simeq 21$ at $\sqrt{s}=M_Z$. The
$\chi$-dependent isospin correction is further suppressed in the
back-to-back region, where the two-jet configuration is dominated by a
single quark flavour in each hemisphere and the charged-to-neutral energy
ratio approaches a constant.

We have verified this picture quantitatively by means of PYTHIA~8.3~\cite{Bierlich:2022pfr} 
simulations of $e^+e^- \to \mathrm{hadrons}$ at $\sqrt{s} = M_Z$ including
NP hadronization effects. The ratio 
$\frac{1}{\sigma^{ch}_{\rm tot}} \frac{d\Sigma^{\rm ch}}{d\cos\chi} / \frac{1}{\sigma_{\rm tot}} \frac{d\Sigma}{d\cos\chi}$ 
is found to be flat in $\chi$ at $1\%$ level across the entire fit 
range $\chi \in [1.8, 3.0]$, increasing to few percent level for $\chi > 3.0$. This confirms that the shape distortion induced 
by the charged-only restriction is well below the theoretical and 
experimental uncertainties of our analysis. 
%The overall normalization 
%factor is $\langle r^2 \rangle \simeq 0.38$, consistent with the known 
%mean charged energy fraction at LEP energies after full hadron decay 
%chains. 
As a further cross-check, we note that our theoretical 
predictions provide an excellent description of the ALEPH data, with a 
$\chi^2$ per degree of freedom of order unity ($\chi^2 =0.94$), and that no tension is 
observed between the ALEPH dataset and the other measurements included in 
the fit. This provides direct empirical evidence that the effect of using 
charged particles only is negligible at the level of precision of the 
present analysis.

\begin{figure}
\centering
\includegraphics[width=0.48\textwidth]{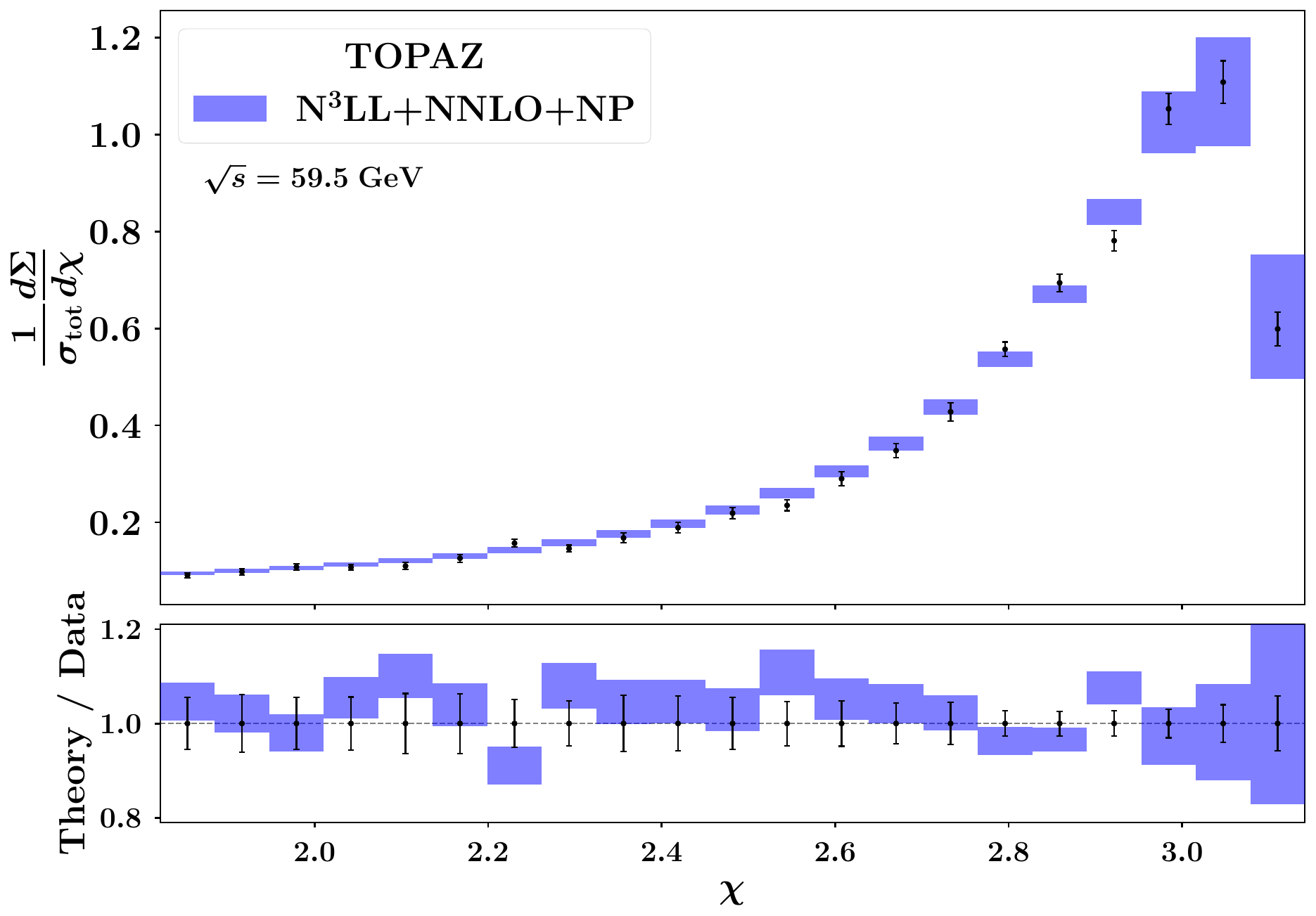}
\includegraphics[width=0.48\textwidth]{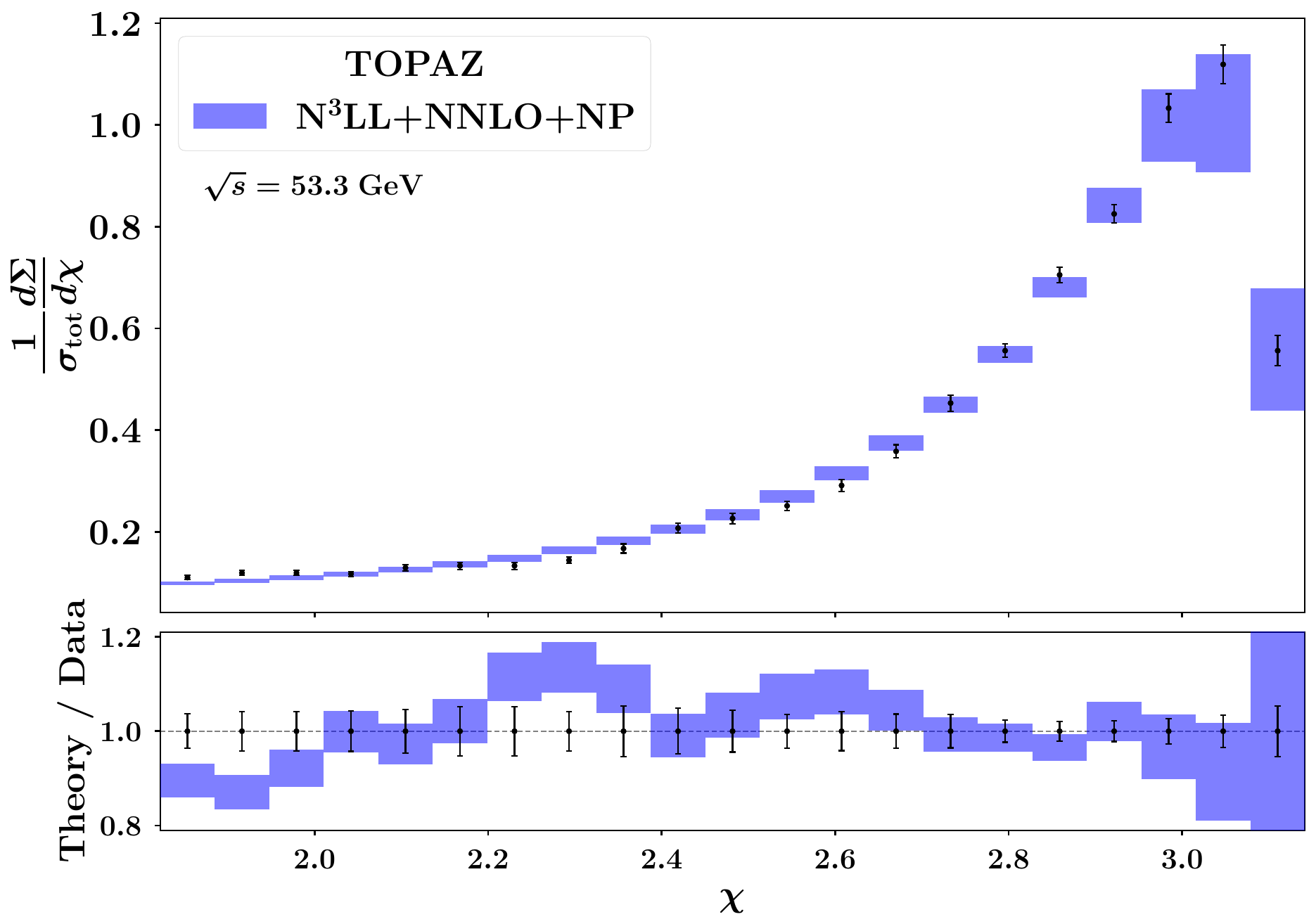}
\caption{Comparison between experimental data and theoretical predictions (blue bands) for the EEC distribution from TOPAZ Coll.\ at $\sqrt{s} = 59.5$ GeV (left) and $\sqrt{s} = 53.3$ GeV (right). The uncertainty bands represent the theoretical uncertainties estimated through scale variations. The lower panels show the ratio of the theoretical results to the experimental data.}
\label{f:TOPAZ}
\end{figure}

\begin{figure}
\centering
\includegraphics[width=0.75\textwidth]{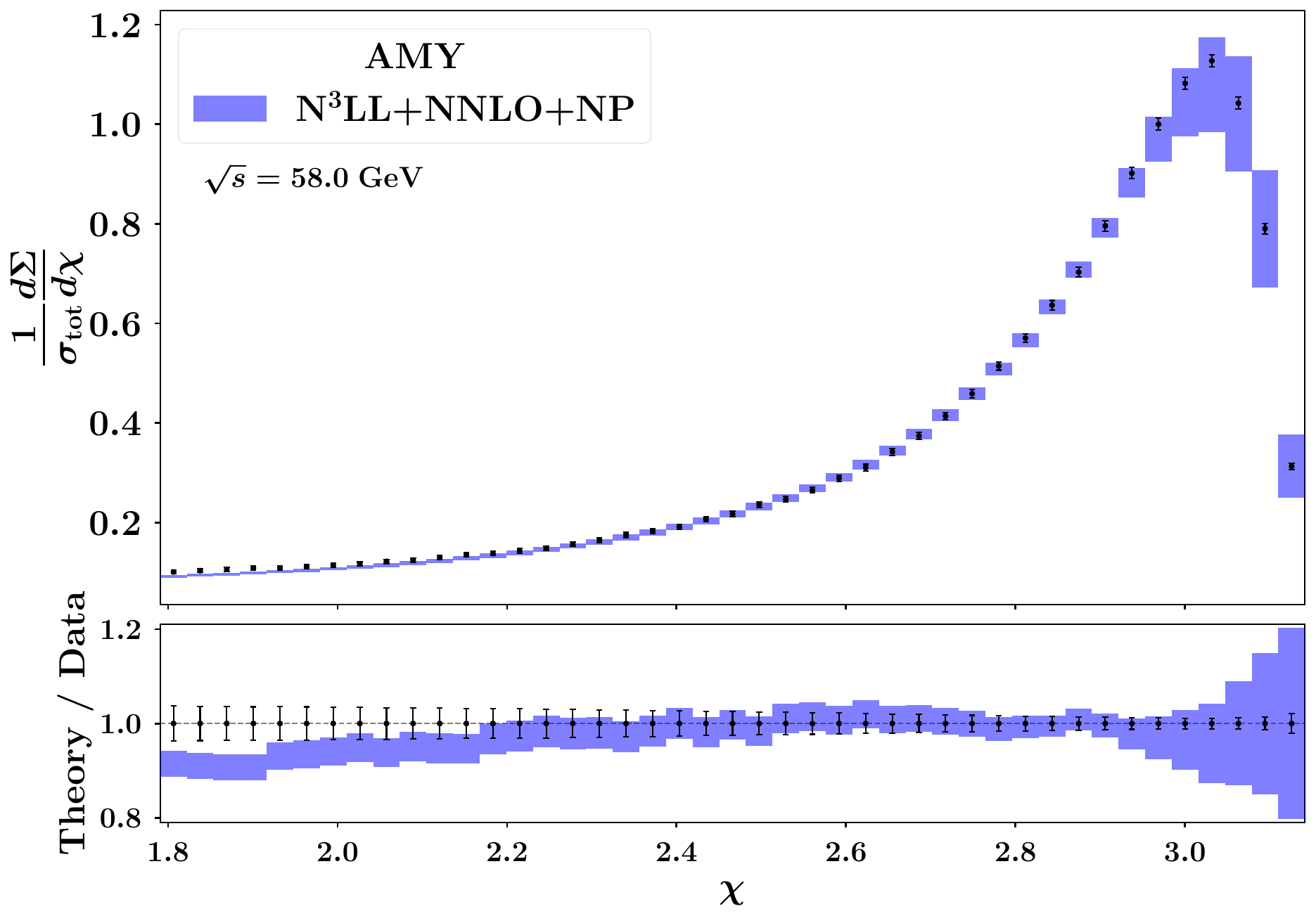}
\caption{Comparison between experimental data and theoretical predictions (blue bands) for the EEC distribution from AMY Coll.\ at $\sqrt{s} = 58.0$ GeV. The uncertainty bands represent the theoretical uncertainties estimated through scale variations. The lower panels show the ratio of the theoretical results to the experimental data.}
\label{f:AMY}
\end{figure}

We now consider data at center-of-mass energies below the $Z$-boson resonance. In Figs.~\ref{f:TOPAZ} and \ref{f:AMY} we compare our theoretical predictions with experimental data from the TOPAZ Collaboration~\cite{TOPAZ:1989yod} at $\sqrt{s} = 59.5$ GeV (left panel) and $\sqrt{s} = 53.3$ GeV (right panel), and from the AMY Collaboration~\cite{AMY:1997waz} at $\sqrt{s} = 58.0$ GeV collected at the TRISTAN (KEK) storage ring. The data are well described within the experimental uncertainties over the entire angular range.

\begin{figure}
\centering
\includegraphics[width=0.49\textwidth]{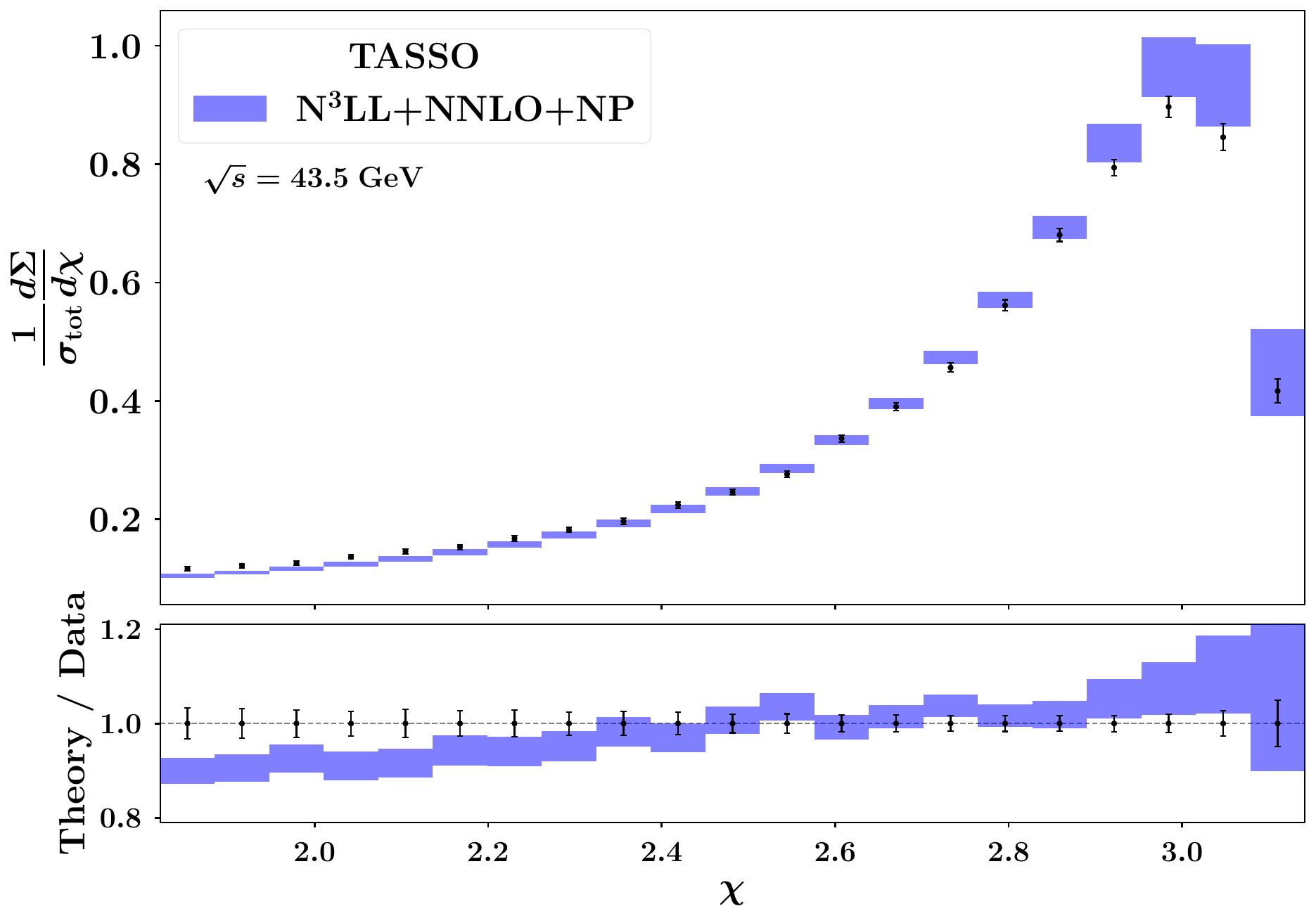}
\includegraphics[width=0.49\textwidth]{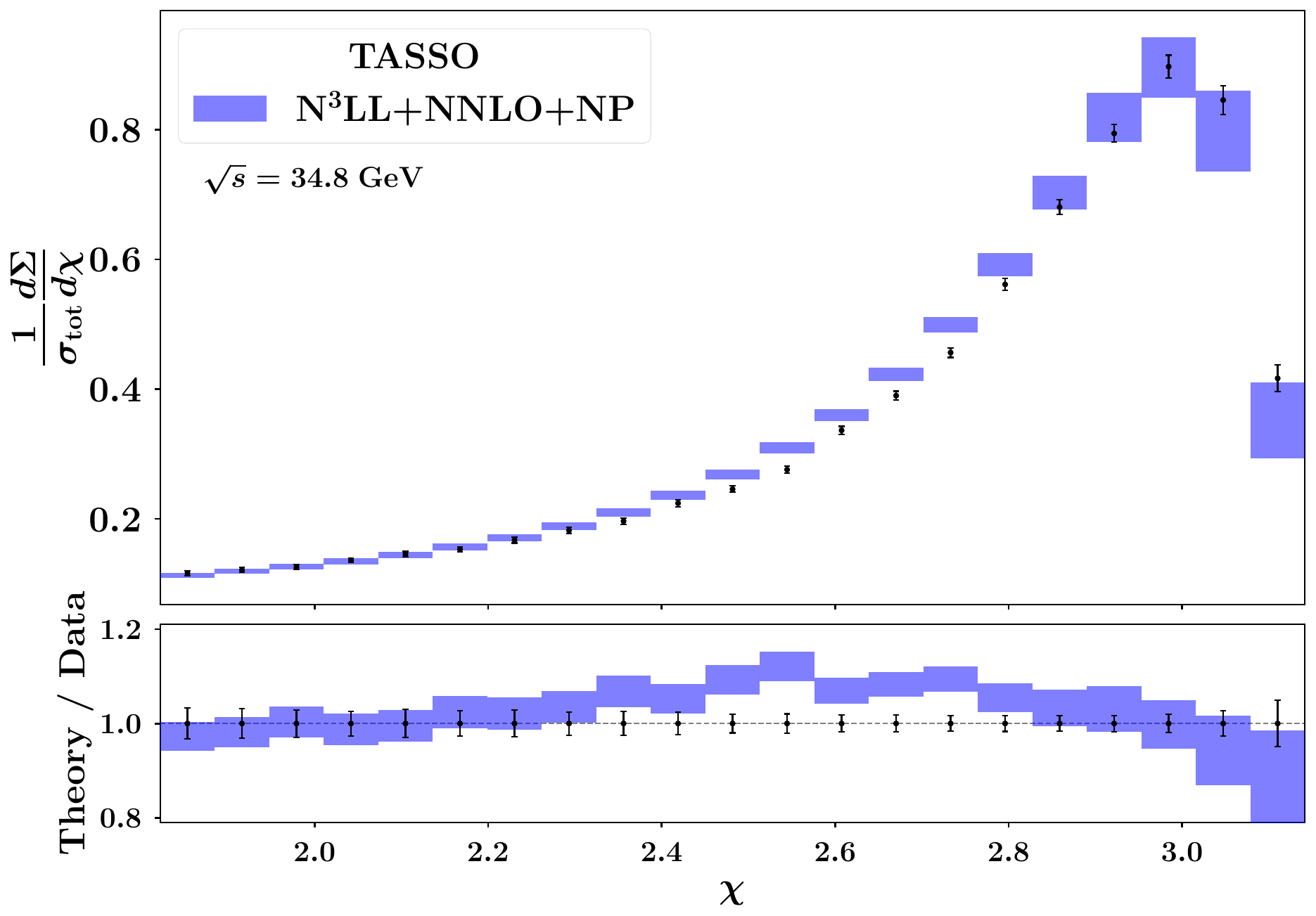}
\includegraphics[width=0.49\textwidth]{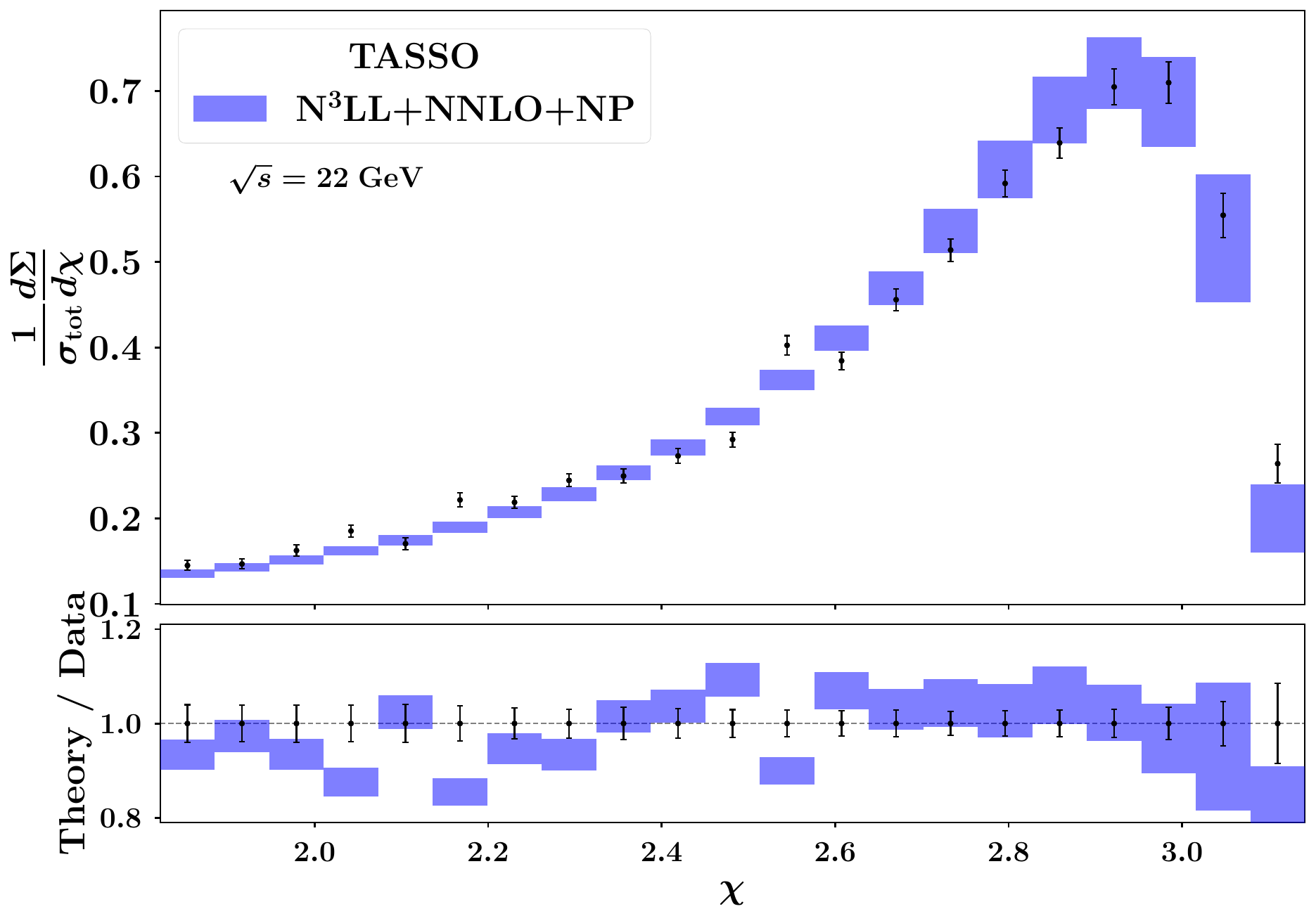}
\includegraphics[width=0.49\textwidth]{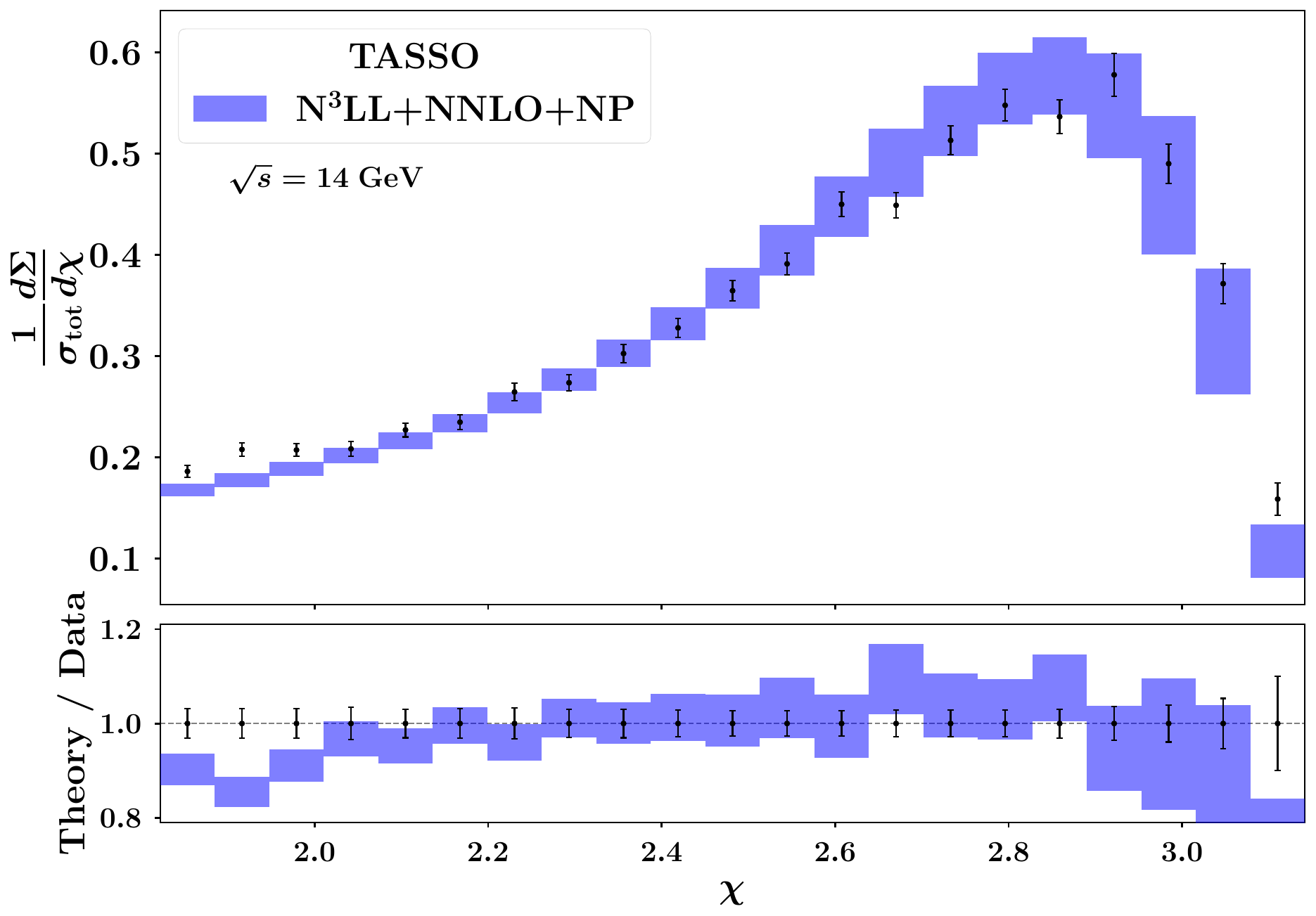}
\caption{Comparison between experimental data and theoretical predictions (blue bands) for the EEC distribution from the TASSO Coll.\ at $\sqrt{s} = 43.5$ GeV (top left), $\sqrt{s} = 34.8$ GeV (top right), $\sqrt{s} = 22$ GeV (bottom left), and $\sqrt{s} = 14$ GeV (bottom right). The uncertainty bands represent the theoretical uncertainties estimated through scale variations. The lower panels show the ratio of the theoretical results to the experimental data.}
\label{f:TASSO}
\end{figure}

\begin{figure}
\centering
\includegraphics[width=0.48\textwidth]{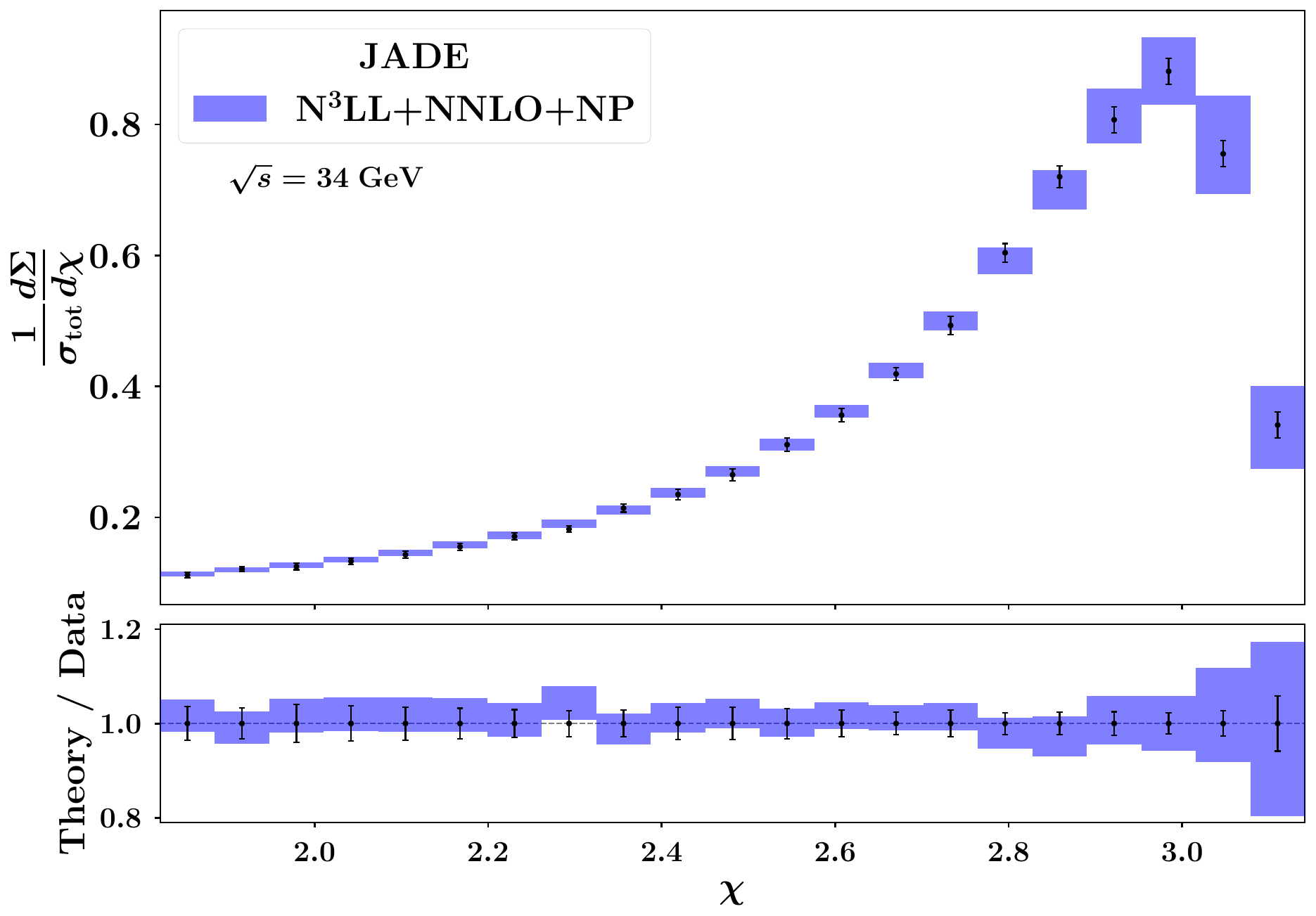}
\includegraphics[width=0.48\textwidth]{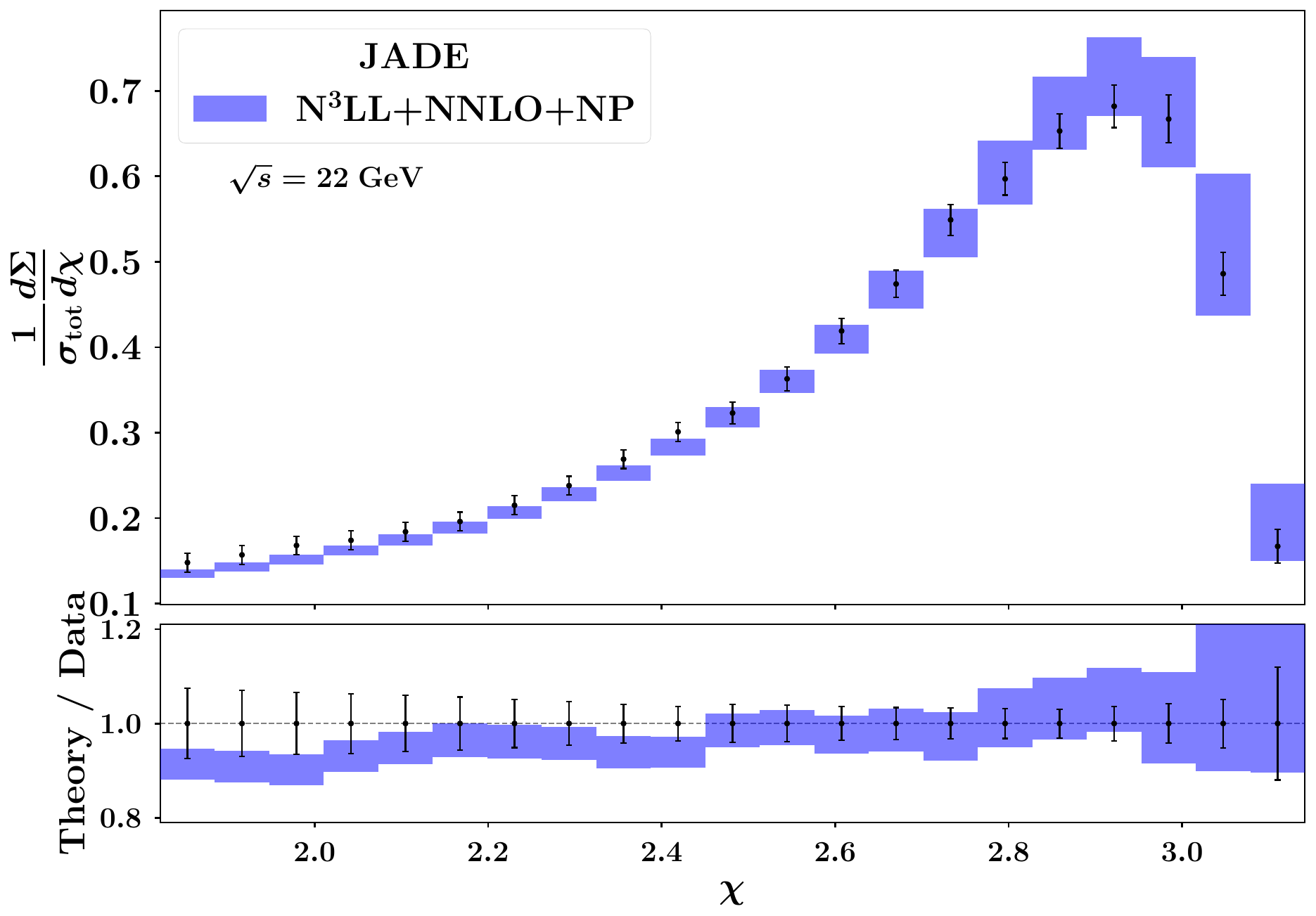}
\includegraphics[width=0.48\textwidth]{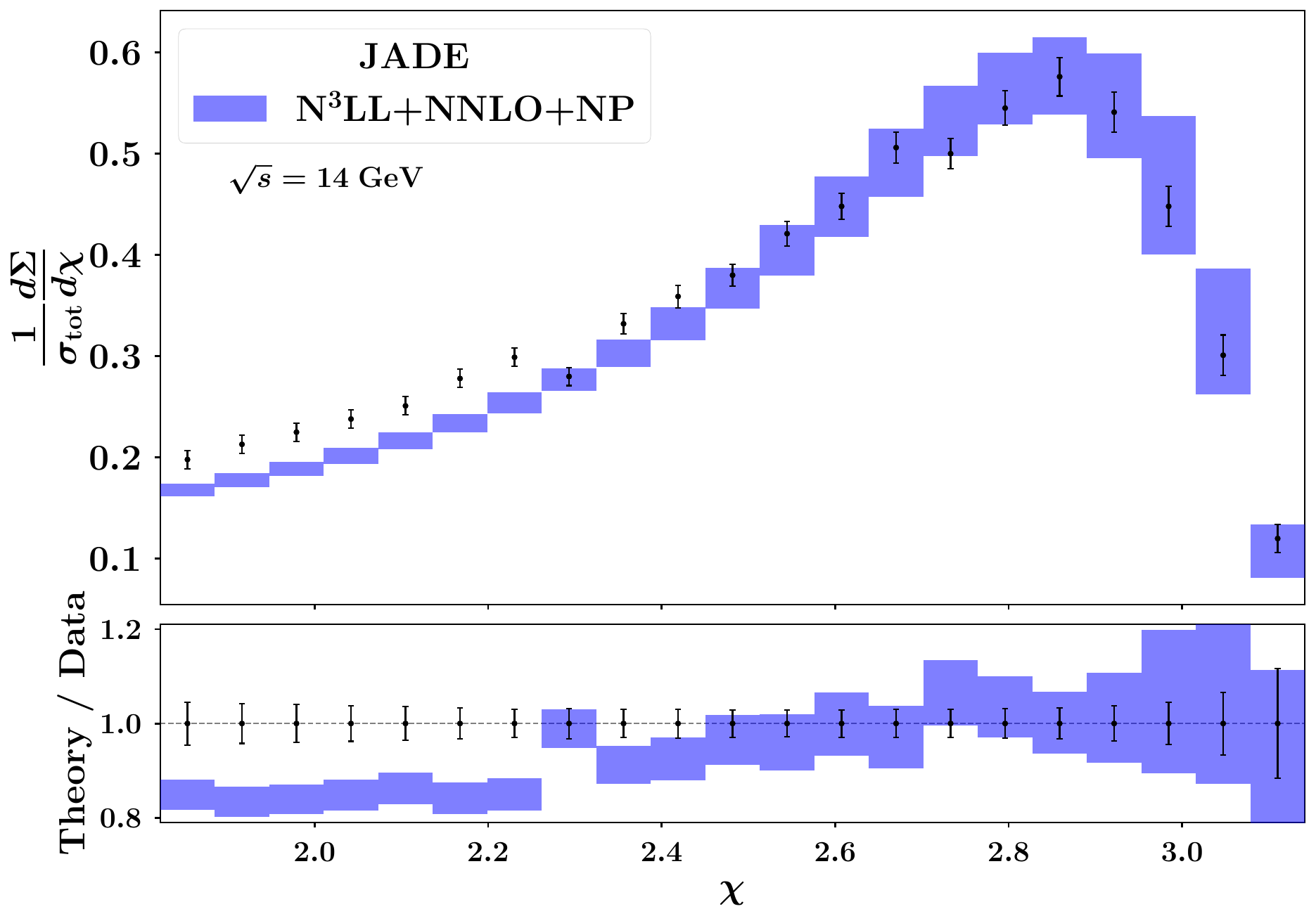}
\includegraphics[width=0.49\textwidth]{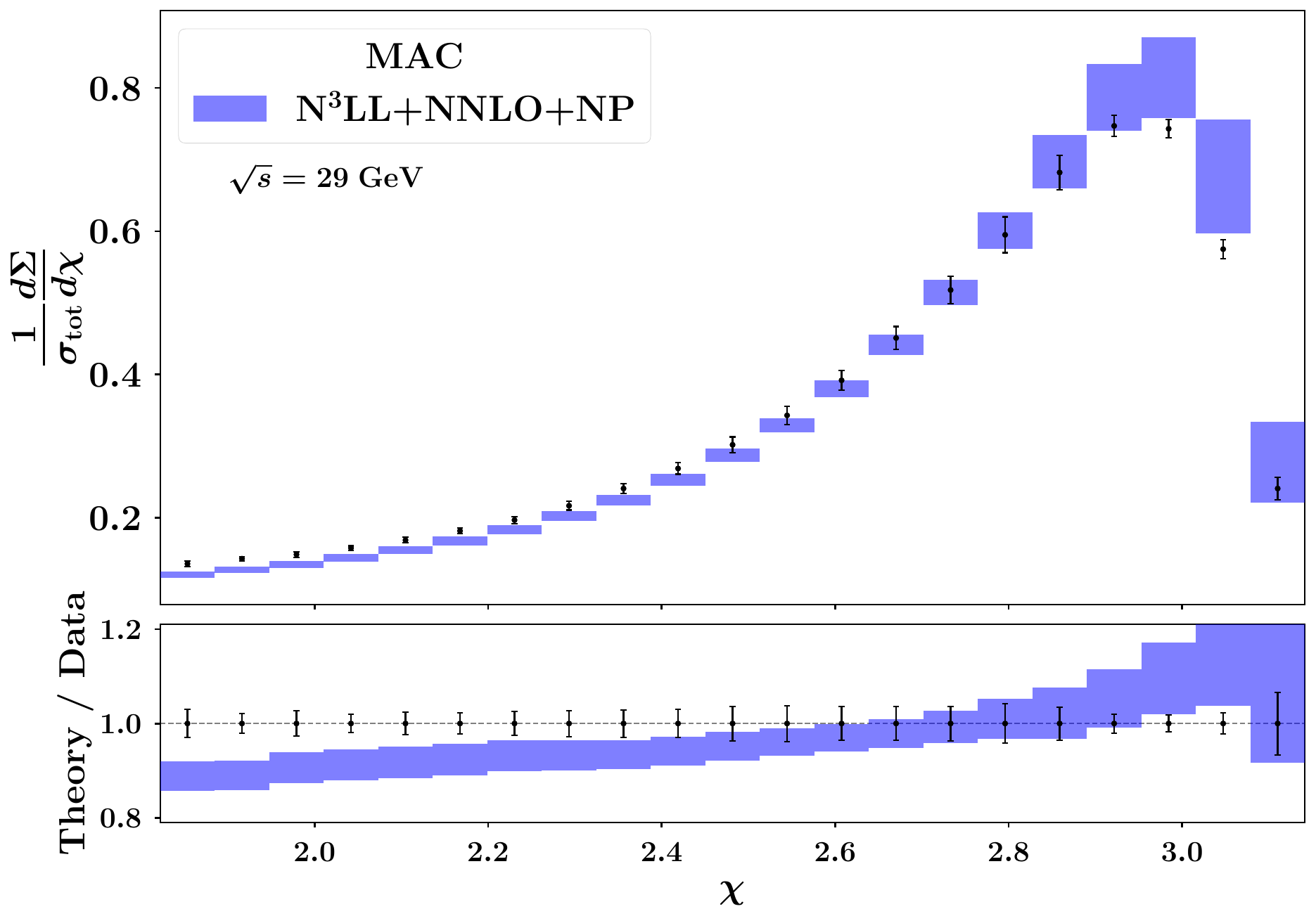}
\caption{Comparison between experimental data and theoretical predictions (blue bands) for the EEC distribution from the JADE Coll. at $\sqrt{s} = 34\,, 22\,, 14$ GeV (top left, top right, bottom left, respectively)
 and MAC Coll. at $\sqrt{s} = 29$ GeV (bottom right). The uncertainty bands represent the theoretical uncertainties estimated through scale variations. 
The lower panels show the ratio of the theoretical results to the experimental data.}
\label{f:JADE}
\end{figure}

We then present, in Figs.~\ref{f:TASSO} and~\ref{f:JADE}, the comparison between our theoretical predictions and the experimental data from the TASSO and JADE Collaborations~\cite{TASSO:1987mcs} at PETRA (DESY), collected at different center-of-mass energies.

The shapes of the experimental distributions are well reproduced by our predictions, providing evidence that the adopted non-perturbative model is sufficiently flexible to describe data across a wide range of energies. As expected, the theoretical uncertainty bands increase at lower energies, reflecting the larger uncertainties associated with higher values of $\alpha_S(Q^2)$ and the relatively larger size of non-perturbative effects.

\begin{figure}
\centering
\includegraphics[width=0.46\textwidth]{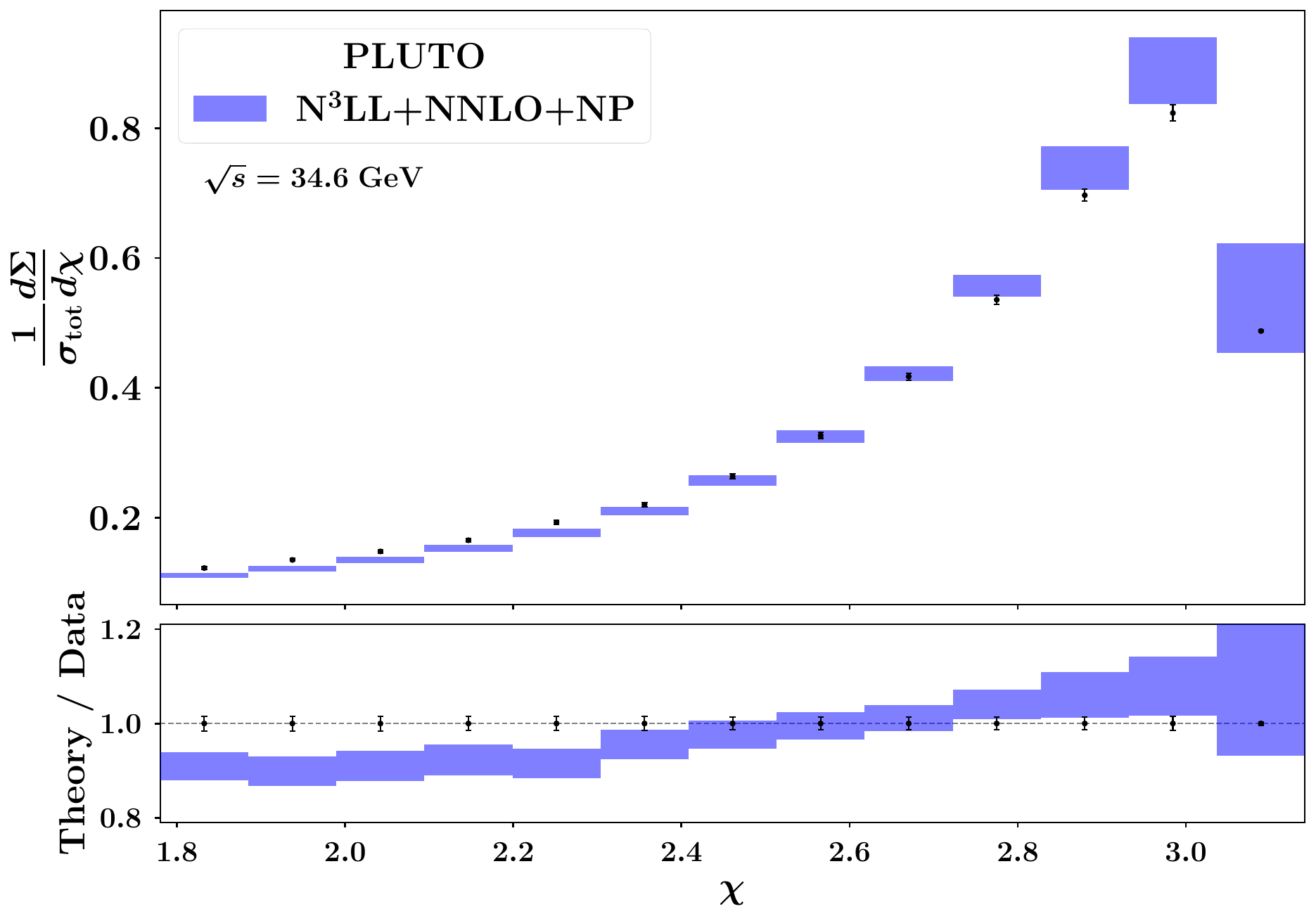}
\includegraphics[width=0.46\textwidth]{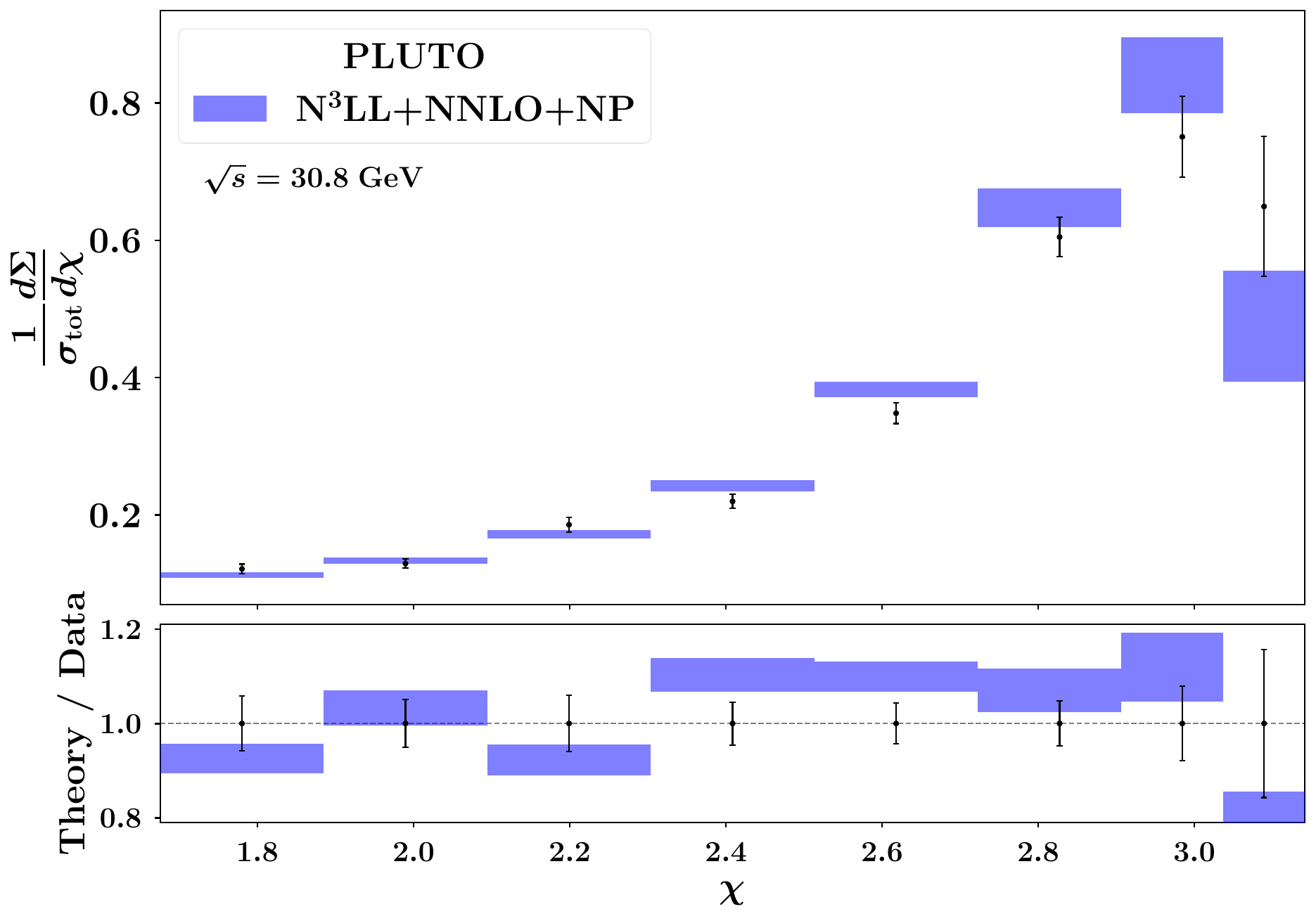}
\includegraphics[width=0.46\textwidth]{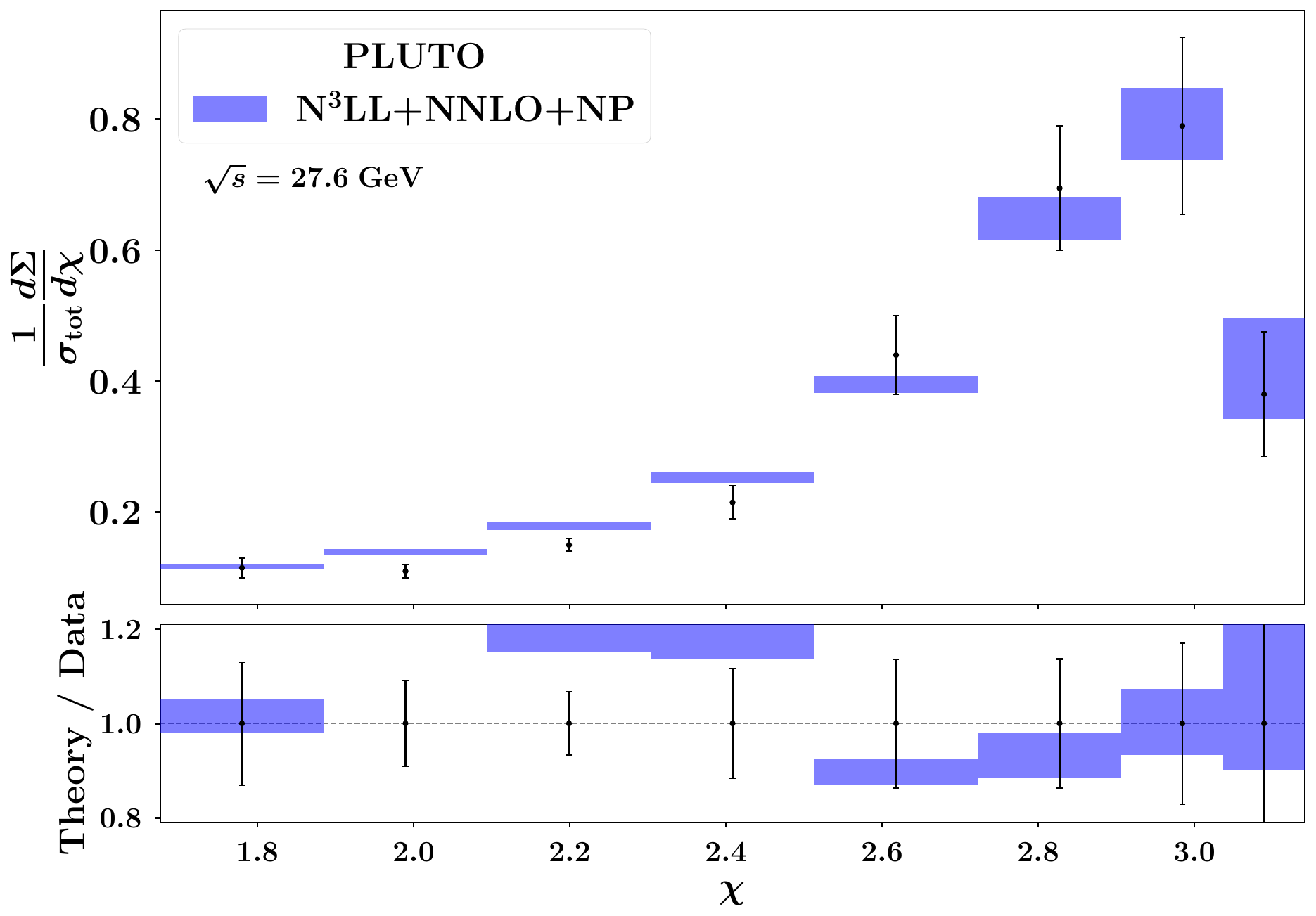}
\includegraphics[width=0.46\textwidth]{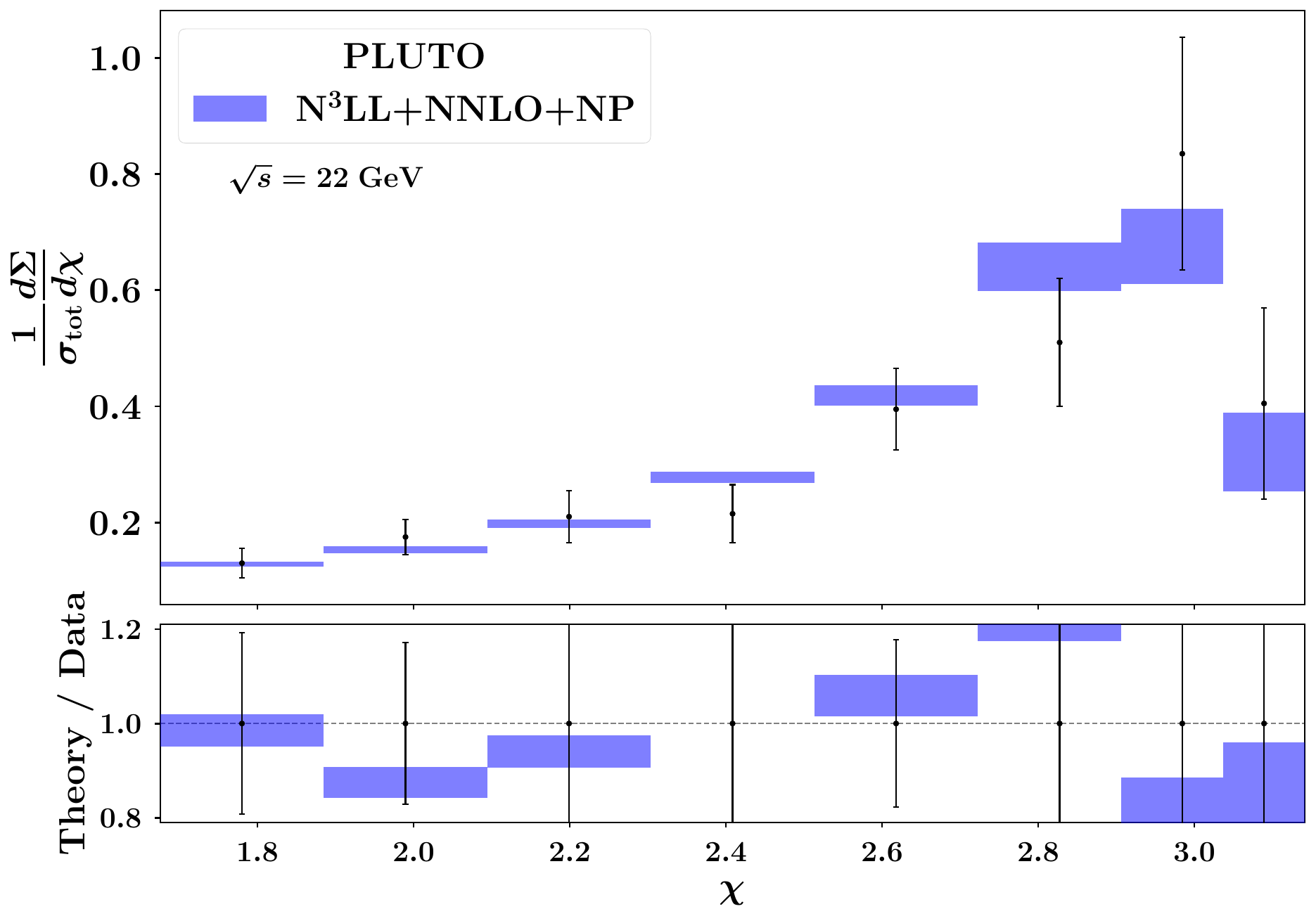}
\includegraphics[width=0.46\textwidth]{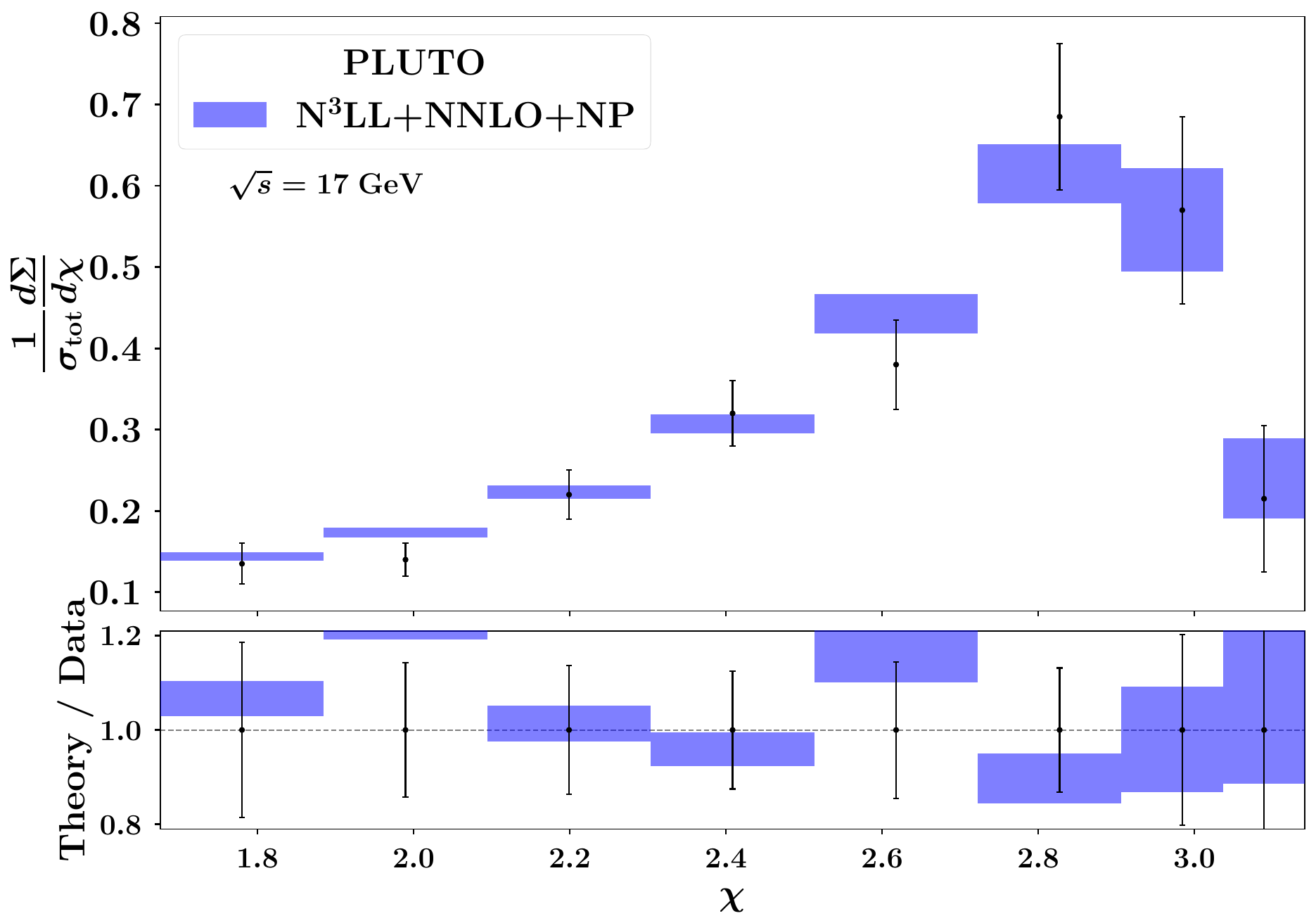}
\includegraphics[width=0.46\textwidth]{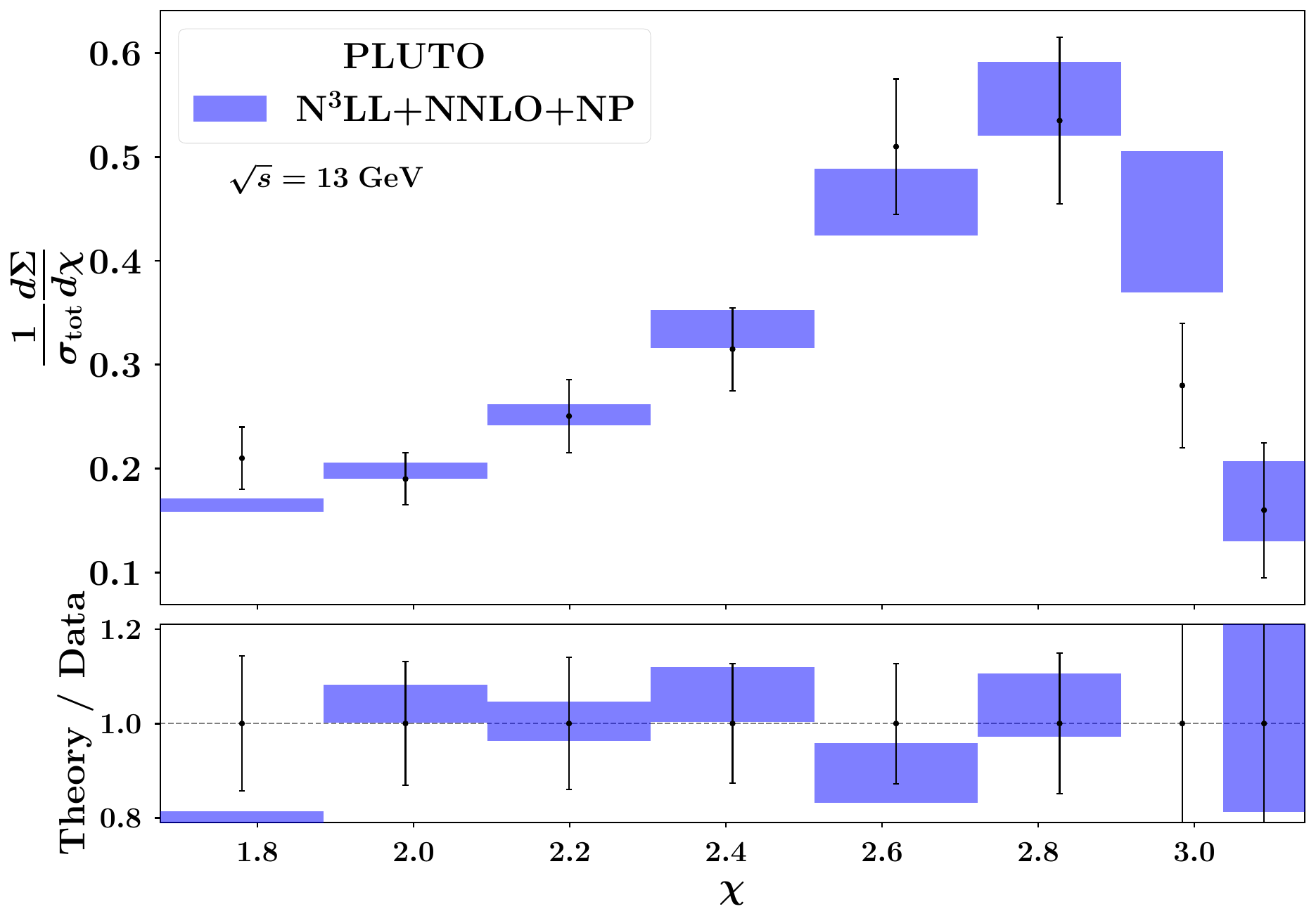}
\includegraphics[width=0.328\textwidth]{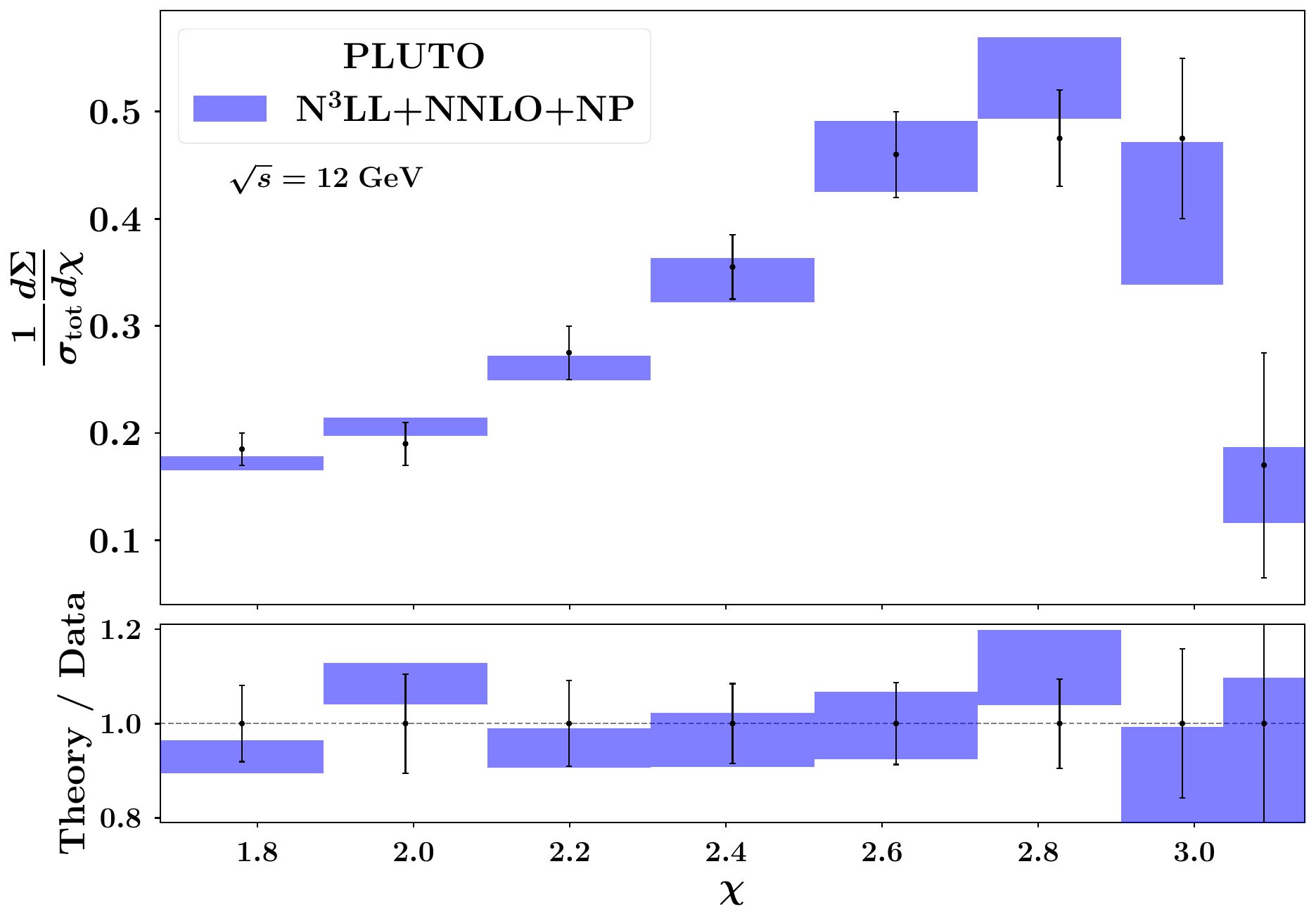}
\includegraphics[width=0.328\textwidth]{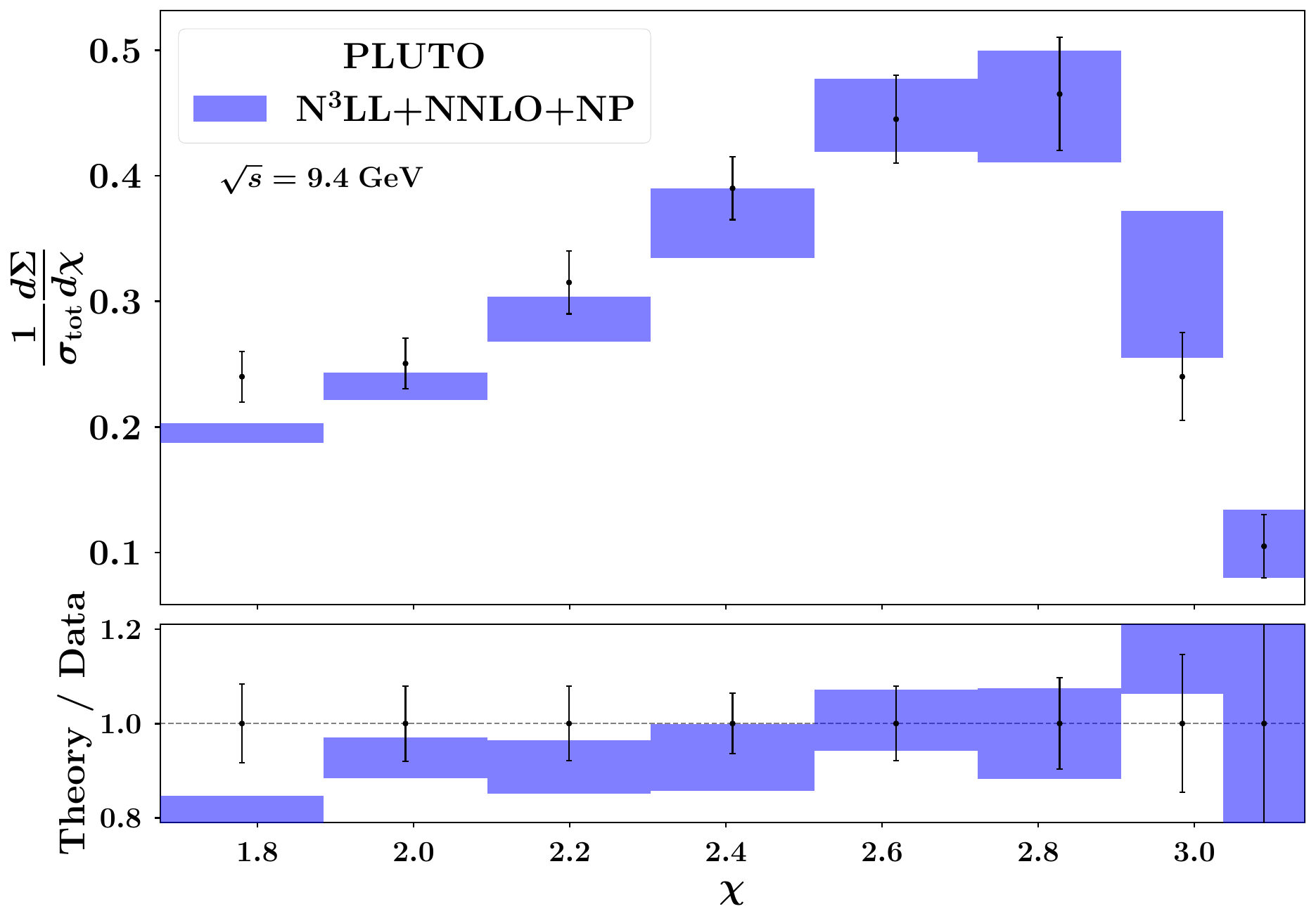}
\includegraphics[width=0.328\textwidth]{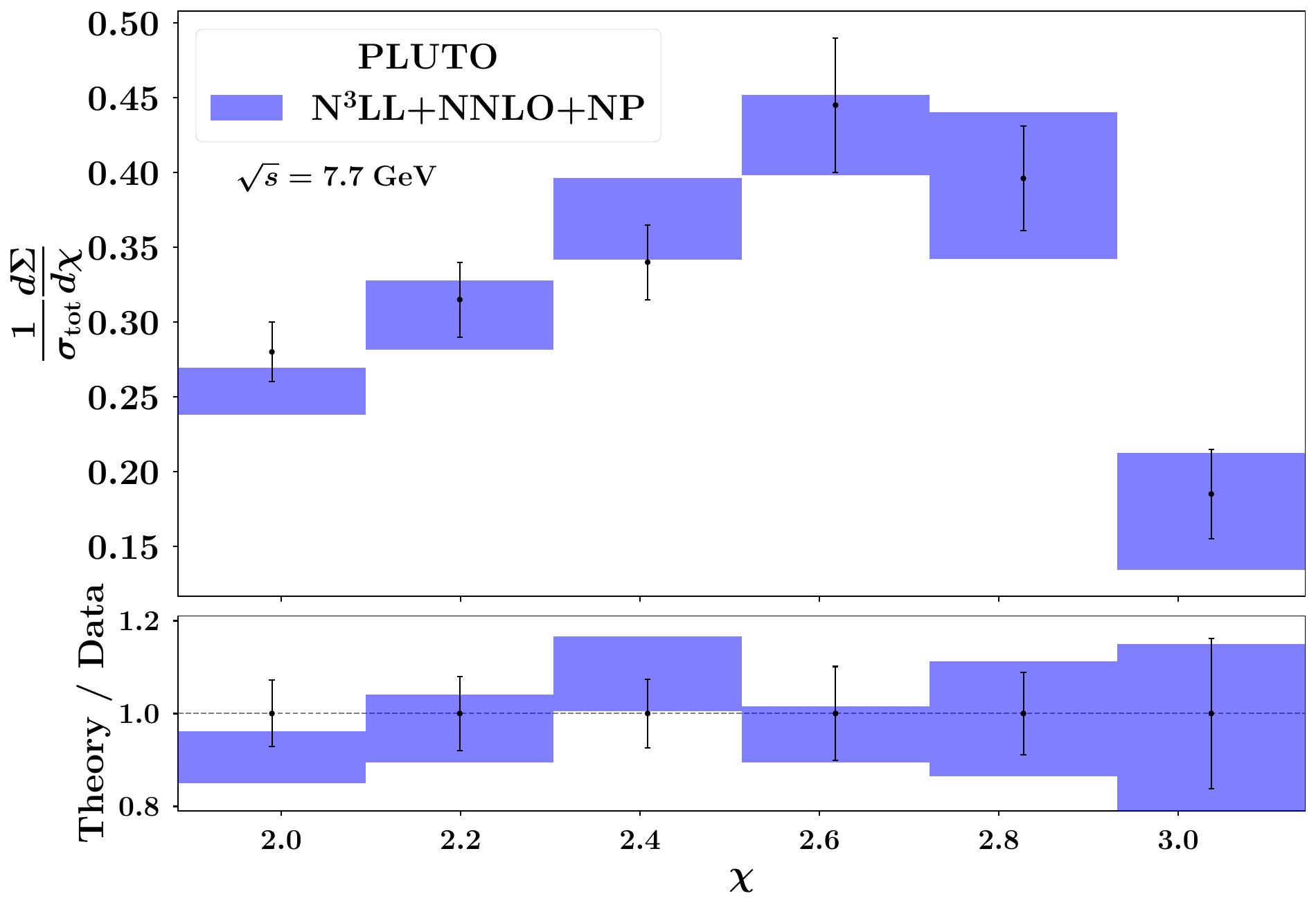}\hspace{-.2cm}
\caption{Comparison between experimental data and theoretical predictions (blue bands) for the EEC distribution from the PLUTO Coll.\ at $\sqrt{s} = 34.6, 30.8, 27.6, 22, 17, 13, 12, 9.4, 7.7$ GeV (from top left to bottom right). The uncertainty bands represent the theoretical uncertainties estimated through scale variations. 
The lower panels show the ratio of the theoretical results to the experimental data.}
\label{f:PLUTO}
\end{figure}

In Fig.~\ref{f:PLUTO} we compare our theoretical predictions with experimental data from the PLUTO Collaboration collected at DORIS II and PETRA (DESY), spanning nine center-of-mass energies from $\sqrt{s} = 34.6\,\mathrm{GeV}$ down to $\sqrt{s} = 7.7\,\mathrm{GeV}$. Despite the relatively large experimental uncertainties of these older datasets, our predictions provide a satisfactory description across the full energy range.

At $\sqrt{s} = 12\,\mathrm{GeV}$ we are sufficiently above the $\Upsilon$ $(b\bar{b})$ resonances to safely neglect their effect. The highest $\Upsilon$ resonance has a mass $m \simeq 11\,\mathrm{GeV}$ and a width $\Gamma \simeq 24\,\mathrm{MeV}$.

At $\sqrt{s} = 9.4\,\text{GeV}$ we are quite close to the $\Upsilon(1S)$ resonance, which has a mass $m=9460\,\mathrm{MeV}$ and a width $\Gamma=54\,\mathrm{MeV}$ (it is the lowest-lying vector resonance, with $J^{PC}=1^{--}$, coupling to the time-like photon coming from $e^+e^-$ annihilation).
The dominant decay channel of this resonance is 
\begin{equation}
\Upsilon(1S) \to 3g,
\end{equation}
i.e.\ three (gluon) jets in the final state.
In a more refined study of the EEC at this energy, one might try to estimate the impact of this resonance.

At $\sqrt{s} = 7.7\,\mathrm{GeV}$ we are below the bottom-quark threshold ($2m_b \sim 10\,\mathrm{GeV}$), so that the number of active flavours is $n_f = 4$. At this energy, charm-mass effects may be non-negligible, since
\begin{equation}
\frac{2m_c}{\sqrt{s}} \sim 0.4.
\end{equation}
Their impact is enhanced by the electric charge of the charm quark,
\begin{equation}
\frac{Q_c^2}{Q_u^2 + Q_d^2 + Q_c^2 + Q_s^2} \simeq 0.4.
\end{equation}

\begin{figure}
\centering
\includegraphics[width=0.48\textwidth]{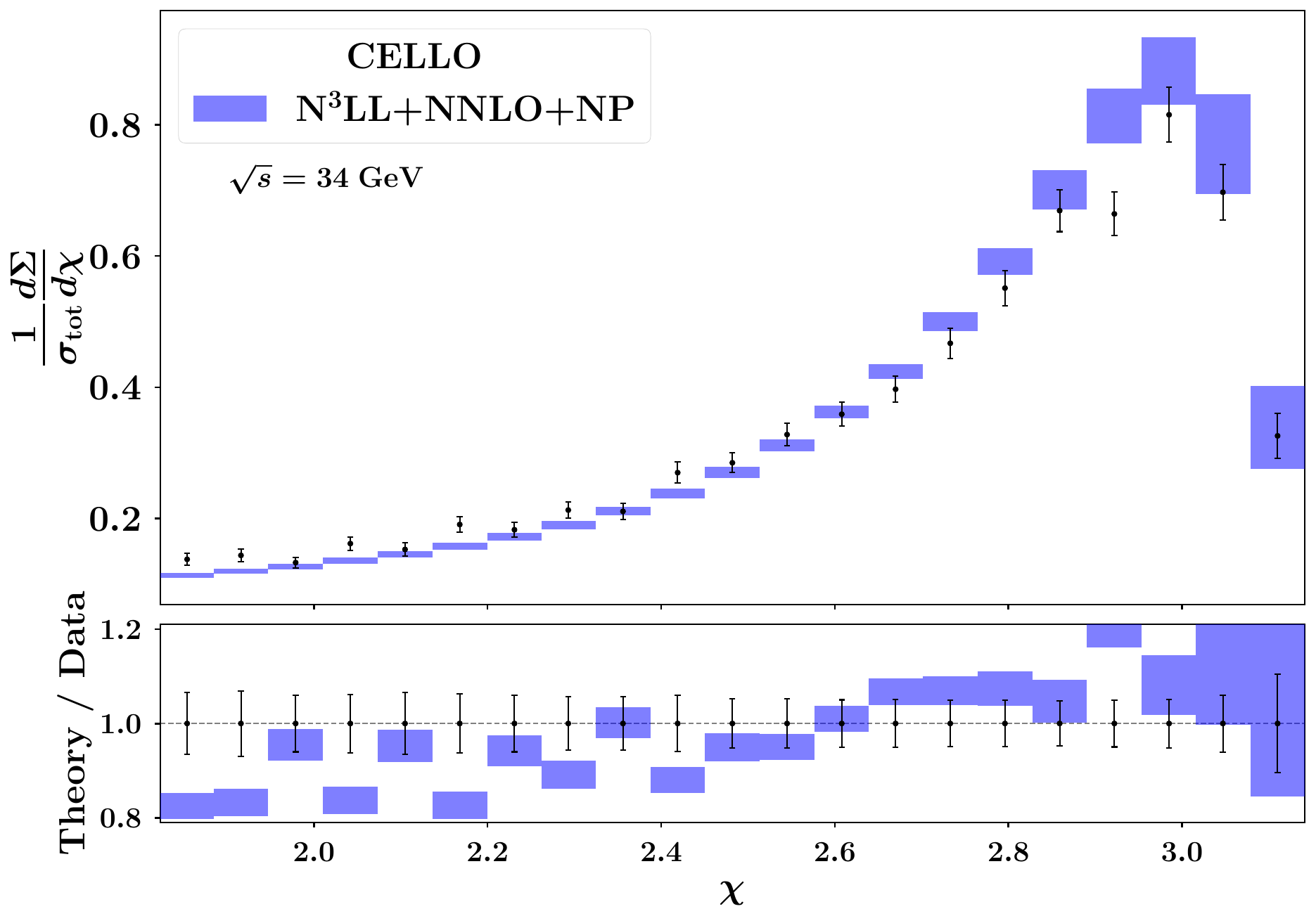}
\includegraphics[width=0.48\textwidth]{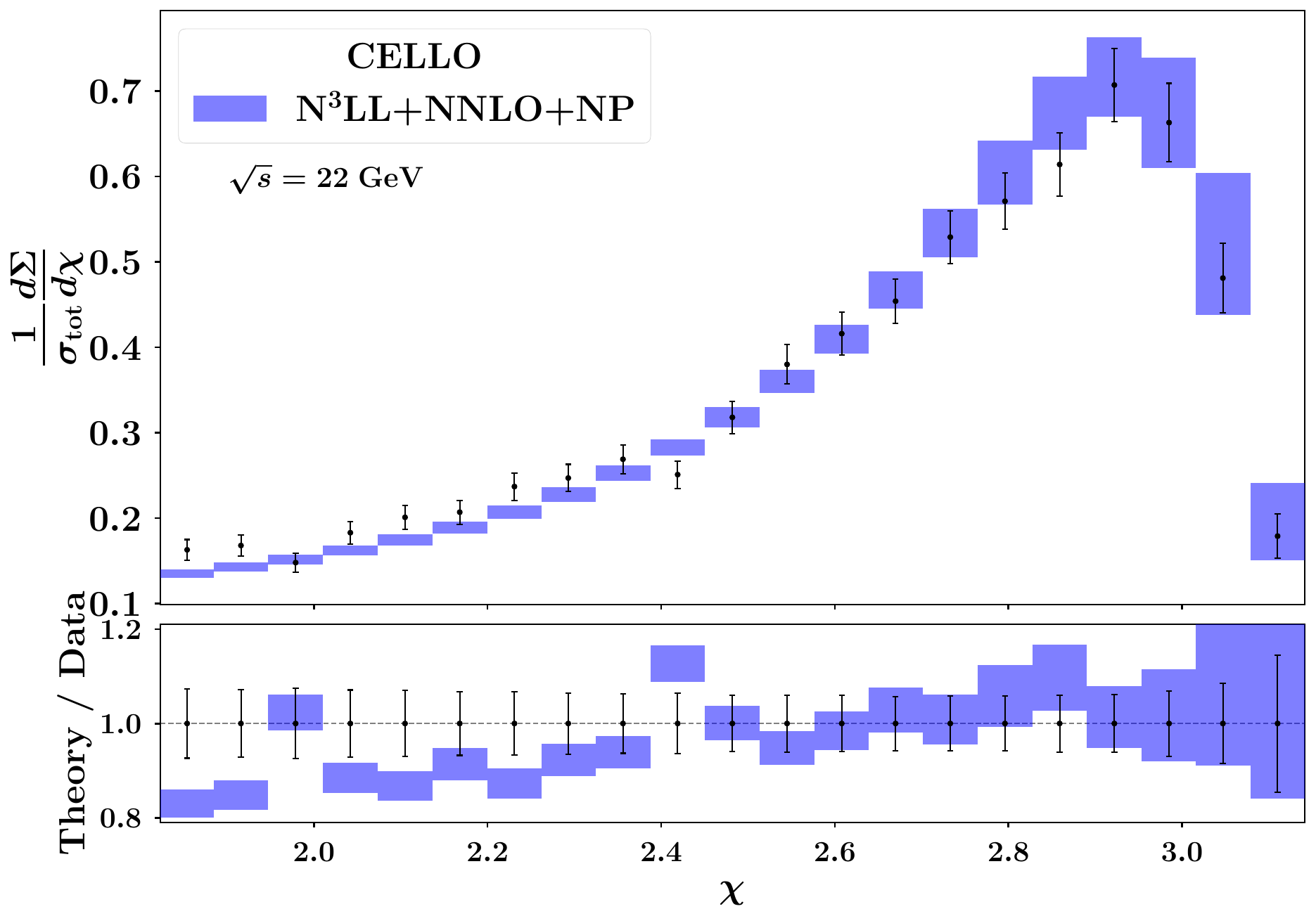}
\caption{Comparison between experimental data and theoretical predictions (blue bands) for the EEC distribution from CELLO Coll. at $\sqrt{s} = 34$ GeV (left) and $\sqrt{s} = 22$ GeV (right). The uncertainty bands represent the theoretical uncertainties estimated through scale variations. The lower panels show the ratio of the theoretical results to the experimental data.}
\label{f:CELLO}
\end{figure}

\begin{figure}
\centering
\includegraphics[width=0.49\textwidth]{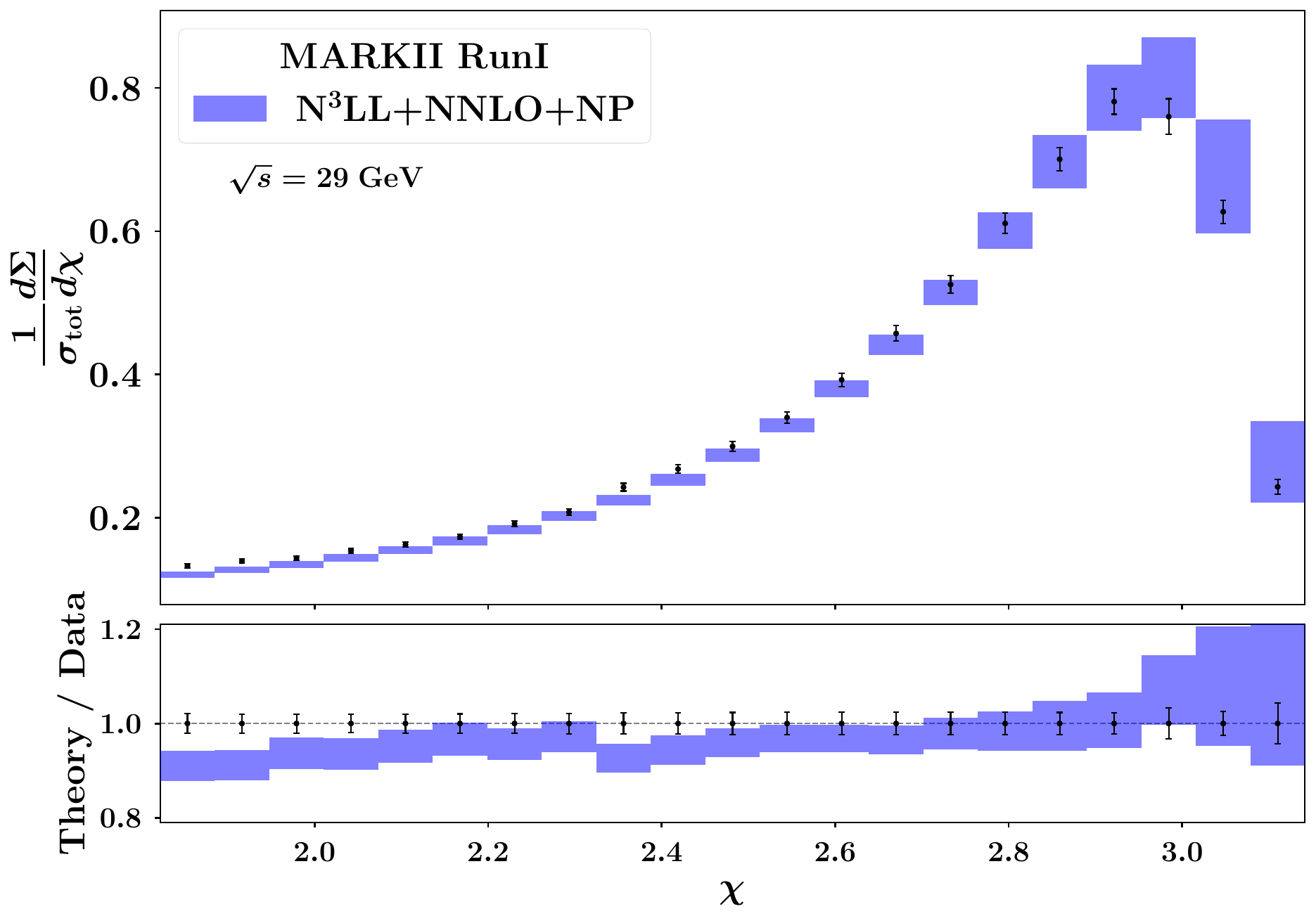}
\includegraphics[width=0.49\textwidth]{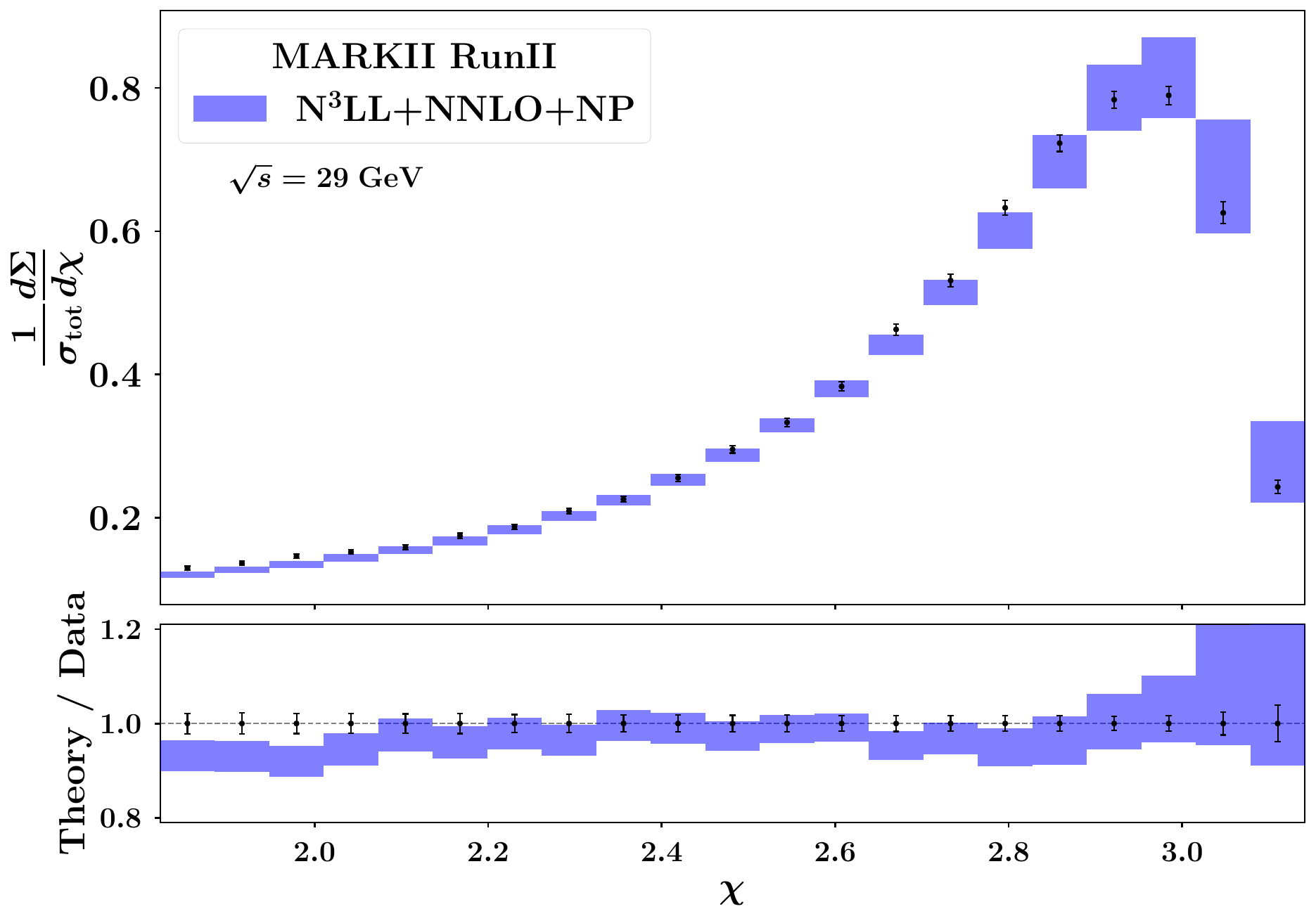}
\caption{Comparison between experimental data and theoretical predictions (blue bands) for the EEC distribution from MARKII Coll. from Run I (left) and Run II (right) $\sqrt{s} = 29$ GeV. The uncertainty bands represent the theoretical uncertainties estimated through scale variations. The lower panels show the ratio of the theoretical results to the experimental data.}
\label{f:MARK}
\end{figure}

In Figs.~\ref{f:CELLO} and~\ref{f:MARK} we compare our theoretical predictions with experimental data from the CELLO Collaboration at $\sqrt{s} = 34.6\,\text{GeV}$ and $\sqrt{s} = 22\,\text{GeV}$ collected at PETRA (DESY), and from the MARKII Collaboration at $\sqrt{s} = 29\,\text{GeV}$ (Run I and Run II) at SLC (SLAC), respectively. Although these datasets exhibit somewhat larger scatter with respect to the theoretical predictions compared to the higher-statistics measurements, an overall satisfactory description is achieved across the full angular range.

Table~\ref{t:par} lists the best-fit values and the associated uncertainties for the fitted parameters, namely $\alpha_S(m_Z^2)$ and the non-perturbative parameters entering Eq.~\eqref{e:SNP}. 
Note that $g_0$ represents the non-perturbative component of the Collins--Soper evolution kernel $g_K(b)$ defined in Eq.~\eqref{e:NP2}, while $f_1$ and $f_2$ characterize the non-perturbative function $f(b)$ defined in Eq.~\eqref{e:NP1}.

The uncertainties are estimated following the procedure described in the previous section and correspond to the total uncertainty, including theoretical, experimental, and correlation-related contributions. In the case of $\alpha_S(m_Z^2)$, we separately indicate in square brackets the three individual contributions to the total uncertainty.

\begin{table}[h]
\vspace{.5cm}    \centering
    \begin{tabular}{|c|c|c|c|c|}
    \hline
       $\alpha_S(m_Z^2)$ & $f_1~[\text{GeV}]$ & $f_2~[\text{GeV}^2]$ & $g_0$ \\
         \hline
          $0.1192 \pm 0.003\,\, [0.0017_{\text{th.}} \pm 0.0015_{\text{exp.}} {\pm0.0016_{\text{corr.}}}] $ & $1.0 \pm 0.3 \, $ & $0.5 \pm 0.3 $ & $0.9 \pm 0.3$ \\
         \hline
    \end{tabular}
\caption{Best-fit values and standard deviations for $\alpha_S(m_Z^2)$ and the non-perturbative parameters.}
\label{t:par}
\end{table}

The extracted value of the strong coupling constant at the $Z^0$ scale, $\alpha_S(m_Z^2)$, is well constrained and consistent with the current PDG average~\cite{ParticleDataGroup:2024cfk}. The fitted non-perturbative parameters are also compatible with those obtained in previous analyses~\cite{Aglietti:2024xwv}, which considered only data at the $Z$-boson mass. We observe that the inclusion of data at $Q < m_Z$ leads to a reduction of the uncertainties for all parameters, in particular for $\alpha_S$.

\begin{figure}
    \centering
    \includegraphics[width=0.5\linewidth]{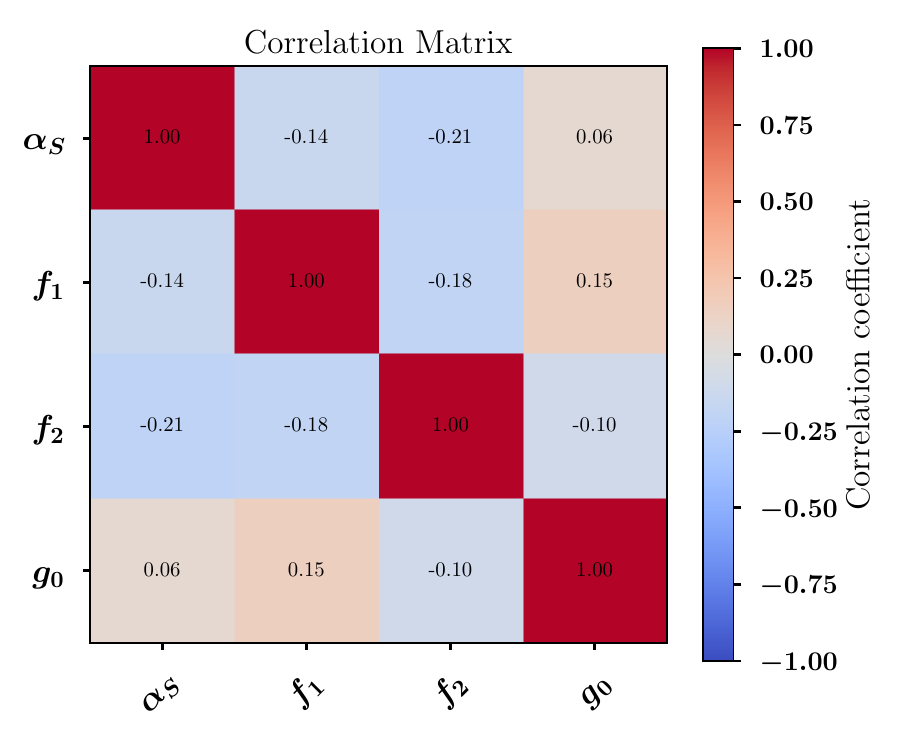}
    \caption{Graphical representation of the correlation matrix for the free fit parameters.}
    \label{f:corr_matrix}
\end{figure}

In Fig.\,\ref{f:corr_matrix} we graphically represent 
the correlation matrix of the (free) fit parameters. 
In general, the absence of strong correlations or anti-correlations between all the
parameters, which never exceed 0.21 in size, 
implies that their determination is not ambiguous.
That is because the fit parameters
play different roles in shaping 
the theoretical distribution.
In particular, while $f_1, f_2$ and $g_0$
are related to power-like effects
in the hard scale,
$\alpha_S$ is related to logarithmic
effects in $Q$.

The (small) anti-correlations
of $f_1$ and $f_2$ with $\alpha_S$
both have the following origin.
By increasing $\alpha_S$, more radiation
is emitted and the Sudakov form factor
becomes broader, with its peak
moving further away from the Born peak
$\chi=\pi$.
That implies that to reproduce
again the data, smaller values of
$f_1$ and $f_2$ are needed.

In order to further assess the impact of the ALEPH dataset on our 
results, and to provide an independent validation of our analysis, we have repeated the global fit 
excluding the ALEPH data entirely, retaining the remaining $610$ data 
points obtaining a $\chi^2 = 1.25$. The extracted parameters are fully consistent with those of the 
baseline fit reported in Table~\ref{t:par}: we find 
$\alpha_S(m_Z^2) = {0.1189} \pm {0.0029}$, with all 
non-perturbative parameters compatible within uncertainties with the 
values in Table~\ref{t:par}. The $\chi^2$ per degree of freedom 
remains of order unity, confirming the stability and robustness of the 
fit with respect to the inclusion or exclusion of the ALEPH dataset. 
As a complementary check, we have also used the parameters extracted 
from the fit \emph{without} ALEPH data to compute a theoretical 
prediction for the ALEPH dataset a posteriori (postdiction). The 
resulting description is excellent, with a $\chi^2$ per degree of freedom 
of order unity ($\chi^2 = 1.02$), and no tension is observed between the theoretical 
prediction and the ALEPH measurements. Taken together, these results 
demonstrate that the ALEPH data are fully compatible with our theoretical 
framework and with the legacy datasets, and that their inclusion does not 
introduce any bias in the determination of $\alpha_S$ or the 
non-perturbative parameters. The ALEPH dataset should therefore be 
regarded as a high-precision independent confirmation of our analysis, 
rather than a source of systematic uncertainty. 
\\
We may therefore conclude that the (resummed) perturbative predictions for the inclusive and charged spectra are the same; in other words, the differences between the two spectra are non-perturbative in nature.

We emphasize that our study demonstrates that analytic dispersive models, such as the DMW model~\cite{Dokshitzer:1999sh}, can provide a consistent description of non-perturbative effects when supplemented by a term with explicit dependence on the hard scale $Q$, as also discussed in Refs.~\cite{Collins:1981va,Collins:1985xx}.
Indeed, as shown by several studies in the literature~\cite{Aglietti:2024xwv,deFlorian:2004mp,Tulipant:2017ybb}, and also confirmed by the present analysis, an analytic NP model (without a $Q$-dependent contribution
as $Q$ is fixed) is capable of describing data at the $Z$-boson resonance
with very good accuracy. 
In this work, we extend the analysis to data at different center-of-mass energies,
finding a very good description of experimental measurements,
once one additional non-perturbative parameter is  added to model the $Q$-dependent evolution. 
In fact, as pointed out in Ref.\,\cite{Kardos:2018kqj}, without the inclusion of a $Q$-dependent non-perturbative term, it would not be possible to achieve a satisfactory description of the experimental data in a global fit. 
We have indeed checked that, by making independent fits for each center-of-mass energy, we obtain fitted values
of the parameters $f_2$ which increase as the energy decreases.
In other words, if we do remove the
NP effects described by the function
$g_K(b)$, and still fit data at different
COM energies, in order to obtain a
good description of the latter, we are forced to take $f_2$ dependent on the energy,
i.e.\ $f_2=f_2(Q)$. 

In general, our results highlight the crucial role played by the $Q$-dependent non perturbative effects in the description of the observable across different center-of-mass energies.

\begin{figure}[h]
\centering
\includegraphics[width=0.9\textwidth]{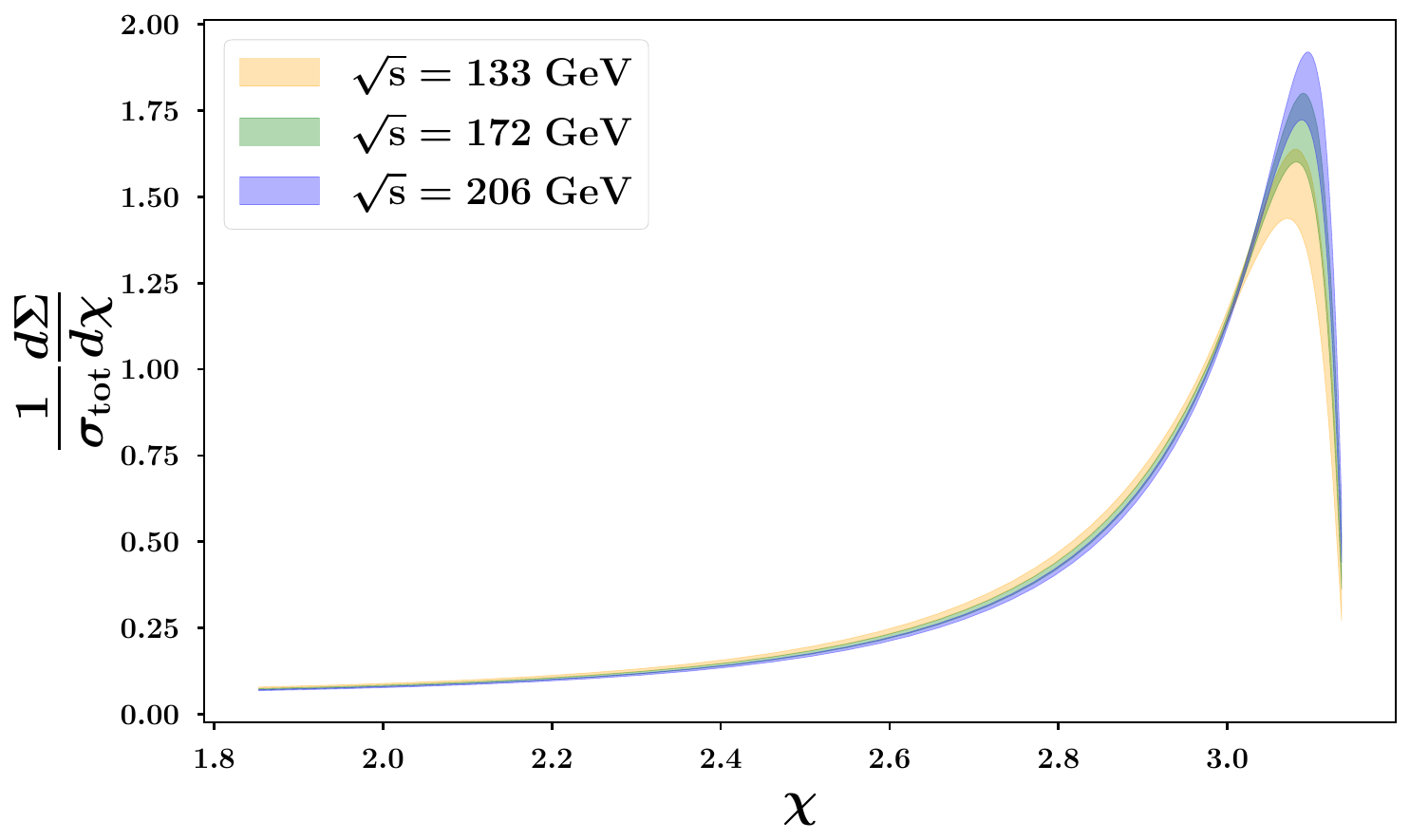}
\caption{Theoretical predictions for the EEC distribution at $\sqrt{s}=133$, $172$, and $206$ GeV.}
\label{f:LEP2}
\end{figure}

In order to explore the potential impact of extending the available measurements to higher center-of-mass energies, we present in 
Fig.\,\ref{f:LEP2} our theoretical predictions for the EEC distribution at $\sqrt{s}=133$, $172$, and $206$~GeV, corresponding to the energy range explored at CERN LEP2. Measurements at these energies would significantly extend the current kinematic coverage and provide a wider lever arm in $Q$. This would allow for a more stringent test of the predicted energy dependence of the observable and improve the determination of the non-perturbative contributions entering the resummed description.

%%%%%%%%%%%%%%%%%%%%%%%%%%%%%%%%%%%
%\subsection{Comparison with ALEPH}
%\LRcom{
%Another novelty of this work is the comparison with the recent reanalysis of archived ALEPH data at the $Z$-boson resonance~\cite{Electron-PositronAlliance:2025wzh,Electron-PositronAlliance:2025fhk}. In Fig.~\ref{f:ALEPH}, we compare our theoretical predictions, obtained from the fit to all other available datasets, with these high-precision measurements.
%\\
%The EEC distribution is shown both as a function of $\chi$ and of $z$ on logarithmic scales. In the $z$ variable, the back-to-back region is strongly compressed and the peak structure is reduced.
%As we can see our postdictions show excellent agreement with the data, with $\chi^2/N_{\text{dat}} = 1.1$. The N$^3$LL+NNLO predictions, supplemented by the analytic dispersive model, provide an accurate description across the full kinematic range.
%\\
%Although the ALEPH observable is not directly comparable to the fully inclusive EEC, we have also performed a fit including these data as a consistency test, finding results fully compatible with our baseline analysis. In this case, we obtain $\chi^2/N_{\text{dat}} = 1.22$ for the global dataset and $\chi^2/N_{\text{dat}} = 1.05$ for the ALEPH data alone. The fitted parameters are fully compatible with those obtained without including ALEPH data.
%}

\subsection{Determination of the  Collins--Soper evolution kernel}
\label{ss:CSK}
 
In recent years, the Collins--Soper kernel has received increasing attention due to its central role in the Transverse Momentum Distributions (TMD) framework, where it controls the evolution of TMDs with respect to the rapidity scale $\zeta$~\cite{Collins:1981uk,Collins:2011zzd}. This interest is further motivated by the possibility of determining the kernel
either from fits to experimental data~\cite{Bacchetta:2024qre,Bacchetta:2025ara,Aslan:2024nqg,Camarda:2025lbt,Moos:2023yfa,Moos:2025sal,Barry:2025glq,Cuerpo:2025zde,Kang:2024dja,Bacchetta:2022awv,Avkhadiev:2025wps}, or theoretically through lattice QCD calculations~\cite{LatticePartonLPC:2023pdv,Avkhadiev:2024mgd,Bollweg:2024zet,Bollweg:2025iol,Tan:2025ofx}. 

Traditionally, the Collins--Soper evolution kernel has been extracted phenomenologically from Semi-Inclusive Deep Inelastic Scattering 
(SIDIS)~\cite{Bacchetta:2024qre,Moos:2025sal} and Drell--Yan (DY) data~\cite{Camarda:2025lbt,Bacchetta:2025ara,Moos:2023yfa,Aslan:2024nqg,Barry:2025glq}. Extracting it from the EEC function~\cite{Cuerpo:2025zde,Kang:2024dja} in 
$e^+e^-$ annihilation provides a highly complementary approach to the former. 
Since 
$e^+e^-$ collisions are free from initial-state hadron effects, our extraction is entirely devoid of uncertainties related to Parton Distribution Functions (PDFs) and initial-state interactions, providing a uniquely clean environment to test the universality of the CS kernel.

The perturbative component of the Collins--Soper kernel satisfies the following
RG evolution equation~\cite{Collins:1984kg}:
\begin{equation}
    \frac{dK\left[b^2Q^2,\alpha_S(Q^2)\right]}{d\ln Q^2} = 
    - \, \gamma_K\left[\alpha_S(Q^2)\right]\, ,
\end{equation}
where $\gamma_K(\alpha_S)$ denotes the cusp anomalous dimension. Solving this equation yields the following formula:
\begin{equation}
    K\left[b^2Q^2,\alpha_S(Q^2)\right] = 
    K\left[b_0^2,\alpha_S(b_0^2/b^2)\right] - \int_{b_0^2/b^2}^{Q^2} \frac{dq^2}{q^2} \gamma_K\left[ \alpha_S(q^2) \right]\,.
\end{equation}
The kernel can be perturbatively expanded as
\begin{align}
K(b^2Q^2,\alpha_S) &= \sum_{n=1}^{\infty} \left( \frac{\alpha_S}{\pi} \right)^{n} K^{(n)}(b^2Q^2) \,.
\end{align}
As shown in Ref.\,\cite{Collins:1984kg} and reported in Ref.\,\cite{Camarda:2025lbt}, the Collins--Soper kernel can be related, within our formalism, to the single and double-logarithmic resummation functions. 

The NP component of the Collins--Soper evolution kernel is given by the energy dependent function $g_K(b)$  
entering the NP Sudakov form factor in Eq.\,\eqref{e:SNP}.
Thus, we have 
\begin{equation}
K(b,Q) = K(b_\star^2 Q^2,\alpha_S) + g_K(b) \, .
\end{equation}

\begin{figure}
    \centering
    \includegraphics[width=0.99\textwidth]{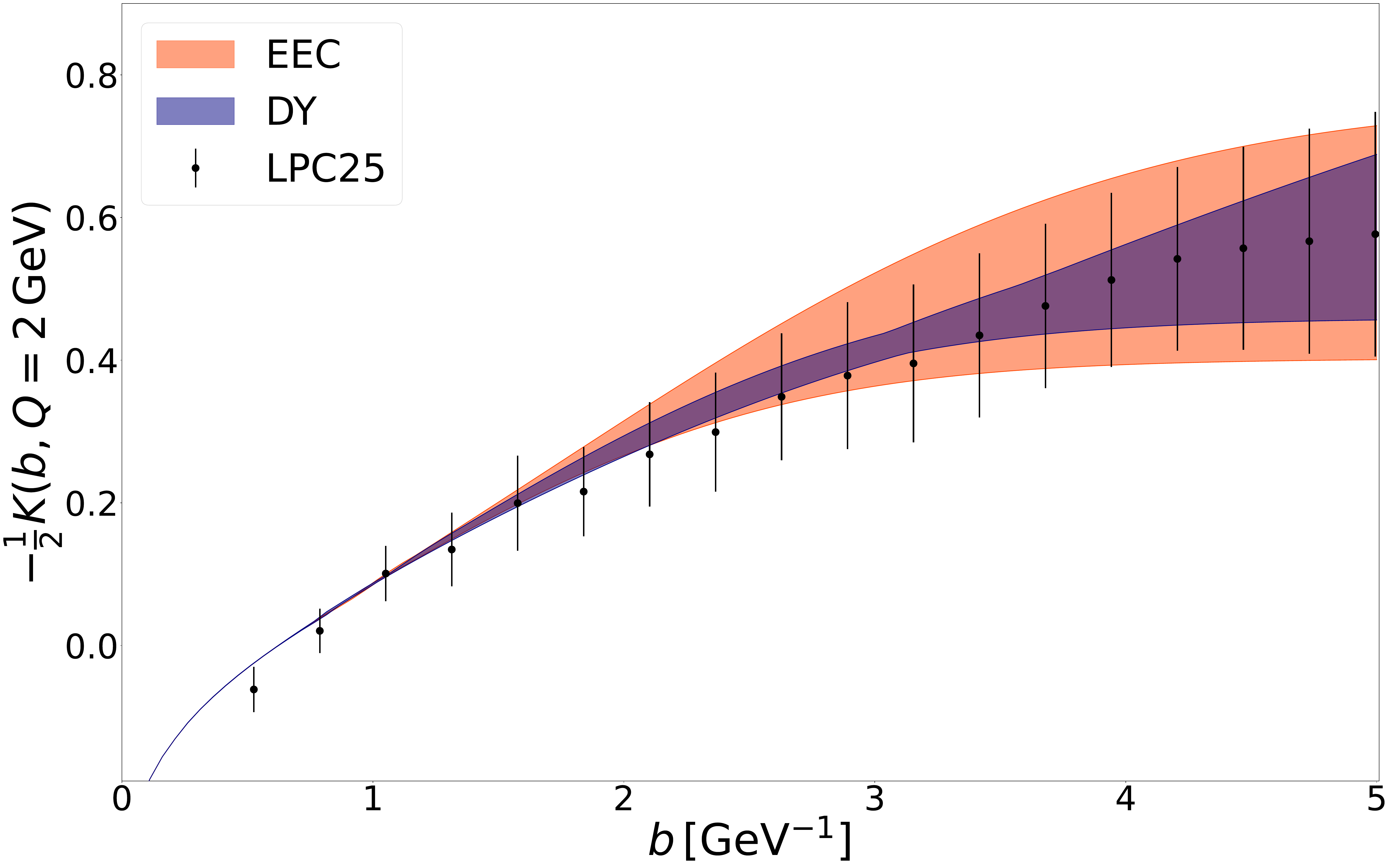}
    \caption{Collins--Soper kernel as a function of impact parameter $b$, at the scale $\mu_R = Q = 2$~GeV, extracted in this work (orange band) and compared with a determination from Drell--Yan data~\cite{Camarda:2025lbt} (blue band) and with the most recent lattice result~\cite{Tan:2025ofx}.}
    \label{f:CSK}
\end{figure}

In Fig.\,\ref{f:CSK} we present the Collins--Soper kernel as a function of the impact parameter $b$ at the hard scale $Q=2\,\text{GeV}$,
extracted from this analysis\footnote{To have a consistent estimation of the Collins--Soper kernel we perform different extractions by varying the value of $b_{\text{max}}$ up to $2.5\,\text{GeV}^{-1}$ as made also in Ref.\,\cite{Camarda:2025lbt}. A similar study was made also in Ref.\,\cite{Cerutti:2026apy}.}, and we compare it with the determination obtained from a global study of Drell--Yan data at different energies~\cite{Camarda:2025lbt}, as well as with the most recent lattice QCD calculation~\cite{Tan:2025ofx}. 
As can be seen, the three determinations are in good mutual agreement for 
$b \lesssim 3 \, \text{GeV}^{-1}$, which corresponds to the region where both the perturbative expansion is reliable and the non-perturbative contributions are well constrained by the data. At higher values of $b$, the uncertainty bands broaden, as expected, given the reduced sensitivity of the kernel to the experimental data in this region. Nonetheless, the overall consistency between our EEC-based extraction, the Drell-Yan determination \cite{Camarda:2025lbt}, and the most recent lattice QCD result~\cite{Tan:2025ofx}, provides a non-trivial test of the universality of the Collins-Soper kernel in different processes and extraction methods.

In addition, we have performed a fit in which the value of $g_0$, i.e.\ the
non-perturbative component of the Collins--Soper kernel, is kept fixed. In this
case we obtain values of $\alpha_S$ and of the two remaining non-perturbative
parameters that are fully compatible with those of our baseline extraction.
This is consistent with what can be inferred from Fig.~\ref{f:corr_matrix},
which displays only weak (anti-)correlations among the parameters: as discussed
below that figure, this is a consequence of the fact that they parametrise
completely different perturbative behaviours.

\subsection{Neglecting heavy-quark mass effects}

The fitted values in Tab.\,\ref{t:par}
have been obtained by including 
bottom-quark mass effects.
In order to reach a general understanding of the form 
and the size of bottom-quark mass effects, 
we have repeated the fits by neglecting
the bottom-quark mass, i.e.\ by setting
$m_b=0$. 
We find that the resulting theoretical predictions, the corresponding values of $\chi^2$
and the extracted value of $\alpha_S(m_Z^2)$ with uncertainties, are comparable to those obtained by including the bottom-quark mass. However, the values of the 
fitted NP parameters are different: 
\begin{equation}
f_1: 1.0 \pm 0.3 \mapsto 0.7 \pm 0.3, 
\quad
f_2: 0.5\pm 0.3 \mapsto 0.8 \pm 0.3, 
\quad
g_0: 0.9 \pm 0.3 \mapsto 1.1 \pm 0.4
\quad
(m_b \mapsto 0).
\nonumber
\end{equation}
That implies that perturbative mass effects can be effectively reabsorbed into a shift of the NP parameters.
However, let us stress that the present analysis, 
involving bottom-quark mass effects,
should be regarded as preliminary, since it 
still uses the massless hard factor and
the massless remainder function
(mass effects are included in the Sudakov
form factor only).
A more refined analysis, involving a wider $\chi$ range, and including {\it all} mass effects
at $\mathcal{O}(\alpha_S)$,
is left to future studies.

\section{Conclusions}
\label{s:conclusions}

In this work, we have carried out a comprehensive global analysis of all available Energy-Energy Correlation (EEC) data in electron-positron  annihilation, spanning center-of-mass energies from approximately 7.7 GeV up to 91.2 GeV, i.e.\ an energy span of over one order of magnitude. 
Our theoretical framework combines next-to-next-to-next-to-leading logarithmic (N$^3$LL) resummation in the back-to-back (two-jet) region with next-to-next-to-leading order (NNLO) fixed-order calculations, consistently matched over the full angular range. 
To compare our (partonic) calculations
with experimental data, we have implemented a non-perturbative model incorporating energy-independent hadronization effects via an analytic dispersive approach, together with an explicit $Q$-dependent term. The latter turns out to be
crucial for a unified description across different energies.

In our analysis, we perform the fit by consistently including data points from both the peak 
and intermediate regions in order to ensure a reliable extraction 
of the non-perturbative parameters and a solid determination of the strong coupling 
constant $\alpha_{S}(m_{Z}^{2})$.
The analysis achieves an excellent description of the complete dataset of 691 data points, with a $\chi^2/N_{\rm d.o.f.}=1.2$, validating both the theoretical framework and the chosen non-perturbative ansatz. As a primary result, we extracted a precise value of the strong coupling constant at the $Z^0$ scale, namely $\alpha_S(m_Z^2) =0.1192 \pm 0.003\,\, [0.0017_{\text{th.}} \pm 0.0015_{\text{exp.}} {\pm0.0016_{\text{corr.}}}] $, where the uncertainties quoted in square brackets indicate, respectively, the theoretical, the experimental uncertainties and the ones related to the correlation-models; the first is estimated through customary scale variations.

A key element of this work is the inclusion of the ALEPH and AMY datasets, which were incorporated into a global EEC fit for the first time. The excellent agreement found across the entire energy range, including these datasets, reinforces the consistency of our theoretical framework and the precision of the extracted value for the strong coupling constant, $\alpha_{S}(m_{Z}^{2})$.
Let us stress that including data over a wide range of center-of-mass energies is crucial, as it allows us to disentangle and determine the energy-dependent component of the non-perturbative model, which is directly related to the Collins--Soper evolution kernel. Thus, our analysis 
enables a novel and accurate extraction of this fundamental kernel from $e^+e^-$ data. 

We have also addressed the fact that the ALEPH dataset is based on 
charged final-state particles only. We have shown, both analytically and 
through dedicated PYTHIA simulations, that this affects the EEC only via 
an approximately shape-independent normalization factor, with negligible 
impact on the extracted value of $\alpha_S(m_Z^2)$ and of the 
non-perturbative parameters.

Finally, we present a preliminary investigation of heavy-quark (bottom-quark) mass effects. 
Including these effects does not significantly 
improve the quality of the fits,
since that basically implies a shift of the non-perturbative parameters. 
In other words, once bottom-quark mass effects
are included and theoretical spectra are refitted,
one finds that the central values
of the non-perturbative parameters
are modified, while the related
uncertainties are not.
This fact implies that a first-principles
(theoretical) determination of the non-perturbative parameters requires a full treatment of heavy quark mass effects, including also those 
ones entering the fixed-order distribution, which is a direction that represents a natural avenue for future work.
\section*{Acknowledgments}
We would like to thank Gabor Somogyi for providing us with the numerical results of Ref.~\cite{Tulipant:2017ybb}, and  Anthony Badea, Hannah Bossi, Francesco Hautmann and Stefan Kluth for useful discussions. We are also grateful to
Dong Young Kim and the members of the AMY Collaboration for providing us the table with the EEC data of Ref.~\cite{AMY:1997waz}

The work of G.F.\ and L.R.\ is partially supported by the Italian Minister of University and Research (MUR) through the research grant 20229KEFAM (PRIN2022, Next Generation EU, CUP H53D23000980006).
\\~\\
\bibliographystyle{JHEP}
\bibliography{biblio}

\end{document}